\documentclass[desactivate]{aa}  
\usepackage{graphicx}
\usepackage{txfonts}
\usepackage{natbib}

\newcommand\un[1]{{\,\rm #1}}
\newcommand\E[1]{\times10^{#1}}
\newcommand\rs[1]{_\mathrm{#1}}
\newcommand\pd[2]{\frac{\partial{#1}}{\partial{#2}}}

\newcommand{\op}[1]{ #1}

\begin{document}
	
\title{Individual particle approach to the diffusive shock acceleration. Effect of the non-uniform flow velocity downstream of the shock}
\titlerunning{DSA with non-uniform flow downstream}
\author{O. Petruk\inst{1,2}
	\and
	T. Kuzyo\inst{2}
}
	
	\institute{INAF - Osservatorio Astronomico di Palermo, Piazza del Parlamento 1, 90134 Palermo, Italy\\
		\email{oleh.petruk@gmail.com}	
		\and
		Institute for Applied Problems in Mechanics and Mathematics, National Academy of Sciences of Ukraine, Naukova St. 3-b, 79060 Lviv, Ukraine
	}
	
	\date{Received ...; accepted ...}
	
\abstract{} 
	{The momentum distribution of particles accelerated at strong non-relativistic shocks may be influenced by the spatial distribution of the flow speed around the shock. This phenomenon becomes evident in the cosmic-ray modified shock, where the particle spectrum itself determines the flow velocity profile upstream. However, what if the flow speed is not uniform downstream as well? Hydrodynamics indicates that its spatial variation over the length scales involved in the acceleration of particles in supernova remnants (SNRs) could be noticeable.} {In the present paper, we address this issue, initially following Bell's approach to particle acceleration and then by solving the kinetic equation. We obtained an analytical solution for the momentum distribution of particles accelerated at the cosmic-ray modified shock with spatially variable flow speed downstream.} {We parameterized the downstream speed profile to illustrate its effect on two model cases, the test particle and non-linear acceleration at the shock.The resulting particle spectrum is generally softer in Sedov SNRs because the flow speed distribution reduces the overall shock compression accessible to particles with higher momenta. On the other hand, the flow structure in young SNRs could lead to harder spectra. The diffusive properties of particles play a crucial role as they determine the distance from the shock, and, as a consequence, the flow speed that particles encounter downstream. We discuss the effect of the plasma velocity gradient to be (partially) responsible for the evolution of the radio index and for the high-energy break visible in gamma rays from some SNRs. \op{We expect that the effect from the gradient of the flow velocity downstream could be prominent in regions of SNRs with higher diffusion coefficient and lower magnetic field, i.e. where acceleration of particles is not very efficient.}}
	{}
	
	
	\maketitle
	
	\section{Introduction}
	
	The theory of diffusive shock acceleration (DSA) was introduced in four independent papers and formulated initially in two alternative approaches. The first approach was through the kinetic equation \citep{1977DoSSR.234.1306K,1977ICRC...11..132A,1978ApJ...221L..29B}, while the second was by using the microscopic description \citep{1978MNRAS.182..147B}. 
	Although Bell's approach is quite intuitive, most of the results related to DSA have been derived by considering the diffusion-convection equation. 
	
	The approach of \cite{1978MNRAS.182..147B} deserves more attention as it may be used to derive the known solutions of the kinetic equation and help with new problems. To illustrate the possibilities of the Bell approach, we obtain in Sect.~\ref{bell2:sectNLA} the solution of \citet{2002APh....16..429B} for the non-linear acceleration (NLA) problem. 
	
	It has been known for a long time that the spectrum of accelerated particles is affected by the non-uniformity of the plasma flow \citep[e.g.][]{1983A&A...122..129B}.  Actually, the NLA solution reflects the non-uniform velocity distribution upstream of the shock, where the particles of higher energies diffuse farther upstream and are scattered back to the shock by scattering centres whose speed differs from the speed of the centres affecting particles of the smaller momenta which are able to penetrate to shorter distances in the pre-shock region. This leads to the formation of a concave particle spectrum \citep[e.g. ][]{2001RPPh...64..429M,2002APh....16..429B} instead of a simple power-law derived from the linear approach where the plasma speed $u$ is considered uniform both upstream and downstream of the shock \citep[e.g. ][]{1983RPPh...46..973D}. 
	
	The common boundary condition for DSA problems is a zero gradient of the flow velocity behind the shock. However, if $du/dx\neq 0$ in the upstream medium affects the particle spectrum shape, what could be the effect if we also consider a non-zero gradient downstream?  
	
	The diffusion length of particles accelerated at the shock may reach  $\sim 0.1$ of the shock radius. At this distance downstream in real objects, e.g. in supernova remnants (SNRs), the flow velocity differs from the immediate post-shock value. Hence, we may expect a `downstream’ effect similar to the `upstream’ one from the non-linear  approach. Moreover, it is known from self-similar solutions \citep[e.g.][]{1982ApJ...258..790C,1959sdmm.book.....S} and hydrodynamic simulations (e.g. recent \citet{2015ApJ...803..101P,2015ApJ...810..168O,2021MNRAS.502.3264G}  for the case of the core-collapse SNe and \citet{2019ApJ...877..136F, 2021ApJ...906...93F} for the type Ia SNe) that the structure of the flow undergoes essential changes during the early evolution of SNRs. Therefore, we may expect that the shape of the spectrum of accelerated particles should vary with time due to the variable non-uniformity of the downstream flow.  
	
	In this paper, we address this effect in the framework of the Bell approach to particle acceleration at shocks. Sect.~\ref{bell2:sectHD} presents detailed motivations by considering the spatial profiles of the plasma velocity behind the SNR shocks from numerical simulations.  Appendix~\ref{bell2:appmovie} includes an animation of the flow speed evolution. 
	In Sect.~\ref{bell2:sectf0}, we use the generalized description of the Bell approach from Sect.~\ref{bell2:sectNLA} to DSA with the non-uniform downstream flow. We consider two practical cases: test-particle and non-linear acceleration (NLA) of particles. Appendix~\ref{bell2:Pderiv} provides relevant details of derivations.
	In Sect.~\ref{bells:dicsus}, we discuss possible scenarios where the effect could be visible. Our conclusions are presented in Sect.~\ref{bell2:conclusions}. Appendix~\ref{bell2:kineq} derives the same solution as in Sect.~\ref{bell2:sectf0} by solving the kinetic equation.

	\section{NLA in the framework of the Bell approach}
	\label{bell2:sectNLA}
	
	Let us consider the shock reference frame with the shock at the coordinate $x=0$. In this frame, the flow velocity of the shock $\mathbf{u}$ is in the positive $x$ direction.
	Particles are accelerated by repeating the acceleration cycle multiple times, i.e. crossing the shock from downstream ($x>0$, index `2') to upstream ($x<0$, index `1') and then again to downstream.  
	Approach of \citet{1978MNRAS.182..147B} consists of finding the increase of the particle momentum per cycle, the probability to participate in the next acceleration cycle and then considering many repetitions. 
	We refer the reader to Sect.~3.3 by \citet{1991SSRv...58..259J} for the extended formulation of the approach. In the present section, we generalize it to the modified shock.
	
	The main idea behind the DSA is that the spatial structure of the flow determines the change in the particle momentum $\dot p\equiv dp/dt$:
	\begin{equation}
		\dot p=-\frac{p}{3}\frac{du}{dx}.
		\label{bell:pdot}
	\end{equation}
	The average momentum gain is derived by integration of this expression,
	\begin{equation}
		\displaystyle\frac{\Delta p}{p}=
		-\frac{1}{3}\int_{x_a}^{x_b}
		\displaystyle\frac{du}{dx}\frac{dx}{v\rs{x}}
		\label{bell:Deltapavegeneral}
	\end{equation}
	where $v\rs{x}=dx/dt$ is the projection of the particle velocity $\mathbf{v}$ on the $x$-axis, and by averaging over the flux. The classic formula 
	\begin{equation}
		\displaystyle\frac{\Delta p}{p}=\displaystyle\frac{4}{v}\frac{u_1-u_2}{3}
		\label{bell:Deltapave}
	\end{equation}
	comes from (\ref{bell:Deltapavegeneral}) for any $x\rs{a}<0$ and $x\rs{b}>0$ if we assume that the upstream $u_1(x)$ and the downstream $u_2(x)$ velocities of the flow are uniform in their domains.
	
	Under the NLA conditions, the pressure of accelerated particles modifies the structure of the flow upstream of the shock, creating some profile $u\rs{1}(x)$. The pressure of these particles downstream is negligible compared to the thermal pressure, so the uniformity of $u\rs{2}(x)=\mathrm{const}$ remains unaffected. 
	
	Let us consider the problem in a simplified and more intuitive way first. We assume that particles with the momentum $p$ can move away from the shock to a maximum distance $|x\rs{p}|$ upstream where the flow speed is $u\rs{p1}=u\rs{1}(x\rs{p})$ \citep{2002APh....16..429B}. Changing $x_a$ to $x\rs{p}$ in (\ref{bell:Deltapavegeneral}), we arrive at the formula (more accurate derivation, refer to Sect.~\ref{bell2:distrfunc})%
	\begin{equation}
		\displaystyle\frac{\Delta p}{p}=\displaystyle\frac{4}{v}\frac{u\rs{p1}-u_2}{3}.
		\label{bell:DeltapaveNLA}
	\end{equation}
	
	In order to participate in the next acceleration cycle, the particle should return to the shock from downstream against the flow direction, i.e. its velocity component $v\rs{x}$ should be $-v\rs{x}>u_2$. Therefore, the ratio of particle fluxes in the positive and the negative $x$ directions gives the probability of continuing acceleration for particles with an isotropic distribution of $\textbf{v}$  \citep{1991SSRv...58..259J}
	\begin{equation}
		P=\displaystyle\left(\frac{1-u_2/v}{1+u_2/v}\right)^2.
		\label{bell2:probdefEJ}
	\end{equation}
	The probability to participate at least in $N$ (or larger) number of cycles is
	\begin{equation}
		P=\displaystyle\prod_{i=0}^{N-1}\left(\frac{1-u_2/v_i}{1+u_2/v_i}\right)^2.
		\label{Bellnla:Probab}
	\end{equation}
	Decomposition of the logarithm of $P$ to the first order on $u_2/v_i$ results in
	\begin{equation}
		\ln P=\displaystyle\sum_{i=0}^{N-1}\frac{4u_2}{v_i}.
		\label{bell:Probdecompos}
	\end{equation}
	
	The original approach generalised the equation for the momentum increase (\ref{bell:DeltapaveNLA}) to $N$ cycles, and its logarithm is decomposed to the first order in $u_2/v_i$. We, instead, express $4/v_i$ from the equation (\ref{bell:DeltapaveNLA}) and substitute (\ref{bell:Probdecompos}) with it. Then, considering that the ratio $\Delta p_i/p_i\sim u_2/c\ll 1$ is small for relativistic particles and the number $N$ is high, we make a transition from the summation to the integration: $\sum \Delta p_i/p_i\rightarrow \int dp'/p'$. After that, we obtain the probability for a particle to be accelerated to momentum $p$ (or higher) from the injection momentum $p\rs{o}$ on the steady background of the nonuniform profile of the flow velocity upstream    
	\begin{equation}
		P(>\!p)=\exp\left(-\int_{p\rs{o}}^{p}\frac{3u_2}{u\rs{p'1}-u_2}\frac{dp'}{p'}\right).
	\end{equation}
	This probability and the definition of 
	the isotropic distribution function $f\rs{o}(p)$ \citep{1991SSRv...58..259J}
	\begin{equation}
		f\rs{o}(p)=-\displaystyle\frac{n}{4\pi p^2}\frac{dP(>\!p)}{dp}
		\label{bell:deffP}
	\end{equation}
	where index `o' refers to the shock location ($x=0$), $n$ is the number density of particles participating in the acceleration process by returning from downstream, 
	results in the same steady-state distribution function of accelerated particles at the modified shock as derived by \cite{2002APh....16..429B} from the alternative approach of solving the kinetic equation, 
	\begin{equation}
		f\rs{o}(p)=\displaystyle\frac{\eta n_{1}}{4\pi p\rs{o}^3}\frac{3u_{1}}{u\rs{p1}-u_2}
		\exp\left[-\int_{p\rs{0}}^{p}\frac{3u\rs{p'1}}{u\rs{p'1}-u_2}\frac{dp'}{p'}\right]
		\label{bell:fnla}
	\end{equation}
	where $\eta$ is the injection efficiency, $n_1$ and $u_1$ should be taken at the same point, e.g. in the immediate pre-shock location (at $x=-0$). 
	It is clear that the shape of the particle spectrum depends on the effective compression factor $\sigma\rs{eff}(p)=u\rs{p1}/u_2$ which is `seen' by particles with momentum $p$. The function  (\ref{bell:fnla}) transforms to the power-law test-particle version by assuming that the flow velocity upstream is uniform, meaning that the same convergence of the flow for particles of any momentum: $\sigma\rs{eff}(p)=u_1/u_2$. The nonlinearity of this solution lies in the fact that $\sigma\rs{eff}(p)$ itself depends on $f(p)$, which requires another equation for these two functions. 
	\cite{2002APh....16..429B} developed a way to relate $\sigma\rs{eff}(p)$ to $f(p)$ by considering the hydrodynamics of the flow. In their notation, $\sigma\rs{eff}(p)=R\rs{tot}U(p)$ where $R\rs{tot}=u_0/u_2$ represents the total compression factor of the shock and $u_0$ is the flow speed far upstream.
	
	\begin{figure*}
		\centering 
		\includegraphics[width=\textwidth]{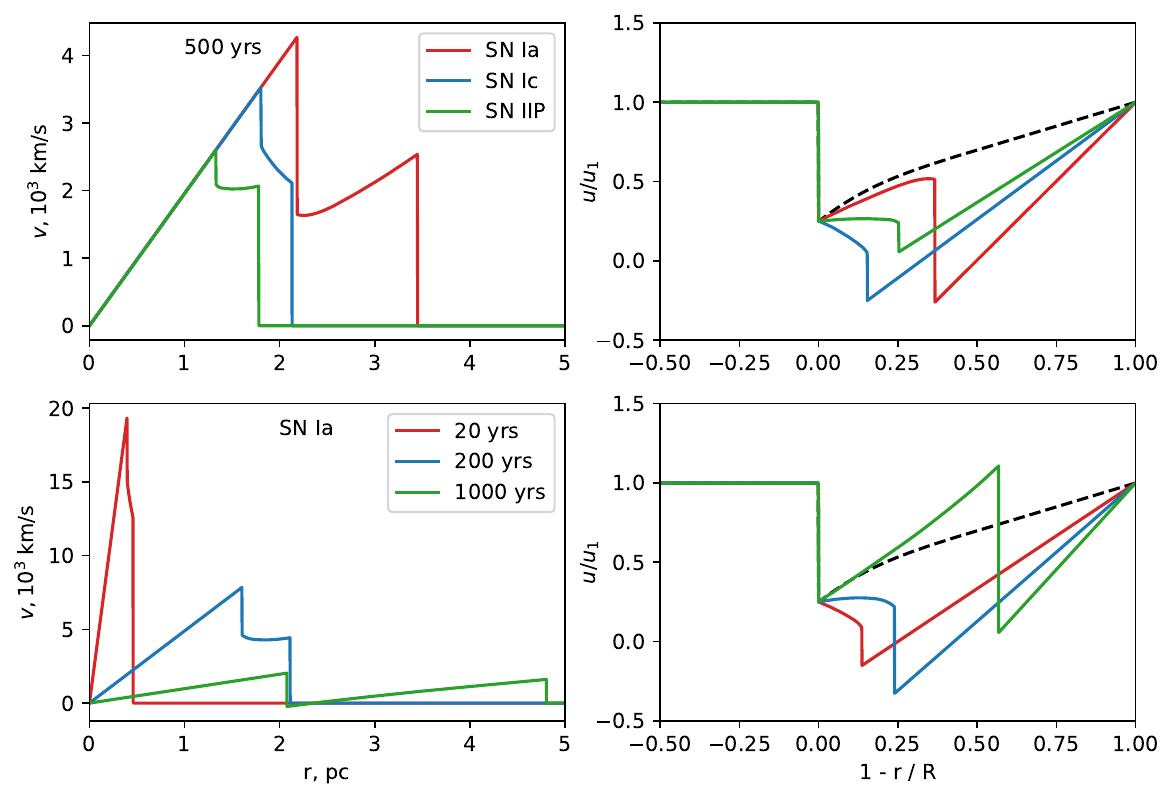}
		\caption{Profiles of the flow velocity $v(r)$ in the observer reference frame (left) and $u(x)$ in the shock reference frame (right) for the three models of SNRs. 
			Upper row: three SNe models at 500 yrs. 
			Lower row: SNIa model at the ages 20, 200, and 1000 yrs. 
			On the plots to the left, the explosion centre is at $r=0$ with the zero velocity $v$ there. 
			On the plots to the right, the horizontal axes display $x=1-r/R$. Therefore, negative values correspond to the upstream region, whereas positive values correspond to the downstream region. The centre of the explosion on the right-hand side plots is at $x=1$, and the velocity $u$ at this point equals the upstream velocity $u_1$.
			We conducted the numerical simulations for the unmodified shocks, therefore $u$ is spatially constant upstream.
			The dashed line represents the velocity profile from the \citet{1959sdmm.book.....S} solution.
		}
		\label{bell2:figshdv}
	\end{figure*}
	
	\section{Profiles of the flow velocity}
	\label{bell2:sectHD}
	
	Studies of the cosmic-ray modified shocks demonstrate that the spatial non-uniformity of the plasma velocity {\it upstream} of the shock affects the shape of the momentum distribution of accelerated particles. It is natural to ask how the {\it downstream} flow structure modifies the particle spectrum. Numerical simulations of CR acceleration which accounts for the complex hydrodynamic structure of the flow inside SNRs demonstrate a variety of shapes of the volume-averaged CR spectra \citep[e.g.][]{2012APh....35..300T,2013A&A...552A.102T,2020A&A...634A..59B}. In the present paper, we would like to consider the effect of the non-uniform flow in the vicinity of a single shock where the CR acceleration happens.  
	
	The length scales involved in the particle acceleration at the forward shock are of the order of the diffusion length which may eventually be up to a few times $0.1R$ where $R$ is the SNR radius. Most of the mass swept up by the forward shock is located right behind the adiabatic shock within about $10\%$ of the radius. The ratio of densities is 
	$\rho(0.9R)/\rho(R)=0.49$ for the Sedov shock in the uniform medium. The variation of the flow velocity in the laboratory reference frame is not so strong, $v(0.9R)/v(R)=0.82$, but it is not negligible.  
	
	The leading motivation for the present study is the fact that the velocity of the flow $u(x)$ is not uniform downstream of the shock on the length scale involved in the particle acceleration. It varies spatially from the value $u_2$ at $x=+0$ in accordance with the profile determined by the hydrodynamics. However, the length scale is small enough to avoid considering the effect of the shock curvature on particle acceleration. In fact, the distance $0.1R$ on the SNR surface corresponds to the sector with a central angle in $\arctan(0.1)=5.7$ degrees. Therefore, we can consider the one-dimensional acceleration problem within such a small sector with a plane shock.
	
	To understand the profiles $u(x)$ we can expect downstream, we suggest referring to our previous paper \citep{2021MNRAS.505..755P}. 
	There, we have investigated the MHD evolution of the remnants of three supernova models, namely, thermonuclear SNIa and two core-collapse SNIc, SNIIP. We have adopted a uniform ISM to model the type SNIa explosion, whereas the core-collapse SNe evolved in a CSM bubble with a density shaped by $\rho\rs{CS} = D r^{-2}$ where the value of $D$ depends on the explosion model. Core-collapse SNe produces a range of explosion conditions, unlike the thermonuclear explosion of a WD star (SNIa event). To account for this, we have selected the two most representative cases of BSG star (SNIc) and RSG star (SNIIP) undergoing the core collapse. The SNIc and SNIIP models differ in the effective explosion mass and CSM number density of a wind-blown bubble. We have performed all numerical simulations with PLUTO MHD code \citep{2007ApJS..170..228M,2012ApJS..198....7M}. The spherically symmetric uniform grid spanned 20 000 zones, providing a very detailed spatial resolution during the entire simulation period. \citet{2021MNRAS.505..755P} provides more details on the technical model setup as well as initial conditions for the explosion models.
	
	Fig.~\ref{bell2:figshdv} shows the spatial distributions of the flow speed for the three models at the age of $500\un{yrs}$ and for three different SNR ages in the case of the SNIa model. The plots in the left column correspond to the laboratory frame, and those in the right column correspond to the shock frame. The velocity profile $u_2(x)$ may rise or decline with the distance from the shock. The distribution evolves with time, e.g. changing from a decreasing to an increasing function of $x$ in the SNIa model (Fig.~\ref{bell2:figshdv} bottom right plot). Animation in Appendix~\ref{bell2:appmovie} shows the detailed evolution of the velocity distributions up to $10\,000\un{yrs}$ in these three models.

	\section{Particle acceleration on the background of the non-uniform flow velocities upstream and downstream}
	\label{bell2:sectf0}
	
	\subsection{Distribution function}
	\label{bell2:distrfunc}
	
	Let us consider the particle acceleration on the flow where the velocity profiles are non-uniform both upstream and downstream of the shock. We start again from the equation (\ref{bell:pdot}).  
	To calculate the increase of momentum per acceleration cycle, we need to spatially separate the integration 
	\begin{equation}
		\Delta p=\int\limits_{-\infty}^{-0} \dot p(x) {\cal X}_1(x)\frac{dx}{v\rs{x}} +\int\limits_{-0}^{+0} \dot p(x){\cal X}_0(x)\frac{dx}{v\rs{x}}+\int\limits_{+0}^{+\infty} \dot p(x){\cal X}_2(x)\frac{dx}{v\rs{x}} 
		\label{bell2:3integrls}
	\end{equation}
	where the functions ${\cal X}_i(x)$ mean of the probability of finding a cosmic ray with a given momentum at a given coordinate $x$. The momentum of a particle is modified not only at the shock transition (where the flow speed changes from $u_1(0)$ to $u_2(0)$) but also in the convergent (or divergent) flows upstream and downstream.
	
	The formula (\ref{bell2:3integrls}) considers one-dimensional flow. We will use it along with the velocity profiles calculated in Sect.~\ref{bell2:sectHD} using the spherical geometry. There are two important factors we have to keep in mind. First, the centre of SNR is at the point $x=R$ in the shock reference frame. Therefore, the upper limit in the last integral should be $R$. However, for the sake of generalisation, we use the notation $+\infty$ since the acceleration region is much smaller than the radius $R$ of SNR. Second, there is a geometrical effect. In the laboratory frame, the integration over the SNR volume is given by $r^Ndr$ where $N=0,1,2$ for the plane, cylindrical, and spherical shocks, respectively. By integrating over $x$ in the shock reference frame, we assume that this effect is accounted for by the probabilities ${\cal X}_i(x)$.
	
	The spatial distribution of particles is given by the functions $f(x,p)$. Therefore, the probability for the accelerated particle to be at a coordinate $x$ is 
	\begin{equation}
		{\cal X}_i(x)=\frac{f_i(x,p)}{f\rs{o}(p)}
		\label{bell3:Xidef}
	\end{equation}
	where $f\rs{o}(p)\equiv f(0,p)$ is the distribution function at the shock.
	Obviously, ${\cal X}_0(x)=1$. 
	
	Equation (\ref{bell:pdot}) reads 
	$\dot p={p}\left(u_1(0)-u_2(0)\right) \delta(x)/3$
	for the second integral in (\ref{bell2:3integrls}). Then, 
	the expression (\ref{bell2:3integrls}) may be written as
	\begin{equation}
		\frac{\Delta p}{p} =
		\frac{1}{3}\frac{u\rs{p1}-u\rs{p2}}{v\rs{x}}
		\label{bell2:Dpp}
	\end{equation}
	with the notations
	\begin{equation}
		u\rs{p1}=u_1(0)-\int_{-\infty}^{-0} {\cal X}_1(x)\frac{du}{dx} dx,
		\label{bell2:defu1p}
	\end{equation}
	\begin{equation}
		u\rs{p2}=u_2(0)-\int_{+\infty}^{+0} {\cal X}_2(x)\frac{du}{dx} dx,
		\label{bell2:defu2p}
	\end{equation}
	The transition to the classical test-particle case is provided by $du/dx=0$. 
	
	The probabilities ${\cal X}_i(x)$ are related to particles that participate in the acceleration around the shock. They do not account for particles escaping upstream or advected downstream. In particular, by equating the diffusive and convective flows in the problem of the test-particle acceleration, we have \citep{1983RPPh...46..973D} 
	\begin{equation}
			{\cal X}_1(x)=\exp\left[\int_{0}^{x}\frac{u_1(x')dx'}{D_1(x',p)}\right]
			\label{bell2:expTPup}
	\end{equation}
		where $D_1$ is the diffusion coefficient.
		For the NLA problem  
		\citep{2005MNRAS.364L..76A,2010APh....33..307C}, the expression for this probability is more complex. However, the same references suggest its accurate approximation:
		\begin{equation}
			{\cal X}_1(x)=\exp\left[\frac{s(p)}{3}\frac{\sigma\rs{s}-1}{\sigma\rs{s}}
			\int_{0}^{x}\frac{u_1(x')dx'}{D_1(x',p)}\right]
		\end{equation}
		where $\sigma\rs{s}=u_1(0)/u_2(0)$ is the sub-shock compression ratio. The probability ${\cal X}_1(x)\simeq \exp(u_1x/D_1)\simeq\exp(-x/x\rs{p})$ where $x\rs{p}\simeq - D_1/u_1$ is the coordinate that corresponds to the diffusion length for particles with momentum $p$. This may roughly be approximated by the Heaviside step function ${\cal X}_1(x)\approx {\cal H}(x-x\rs{p})$. This is why we used $u\rs{p}\approx u_1(x\rs{p})$ in Sect.~\ref{bell2:sectNLA}. 
		
		In the test particle problem, a formula similar to equation (\ref{bell2:expTPup}) may be derived in the same way for the region behind the shock. In the future, a derivation of $f_2(x,p)$ and ${\cal X}_2(x)$ in a more general case, considering NLA effects and $du_2/dx\neq 0$ should be taken into account. In the present paper, we assume that the probability downstream is similar to that upstream, Which means that ${\cal X}_2(x)\simeq\exp(-x/x\rs{p})\approx {\cal H}(x\rs{p}-x)$ and thus $u\rs{p2}\approx u_2(x\rs{p})$. In fact, at a given point $x$ downstream, only a fraction $\xi$ of particles with momentum $p$ can return to the shock and continue acceleration. A fraction $1-\xi$ is advected downstream from this point and never returns to the shock. The fraction $\xi$ decreases with distance from the shock because the particles can return from about one diffusion length. This distance is $x\rs{p}\simeq D_2(p)/u_2\propto p^\alpha/u_2$ with positive $\alpha$, i.e. there are progressively fewer particles with momentum $p$ able to return being farther from the shock.

	The equation (\ref{bell2:Dpp}) gives the average $\Delta p/p$ for the first half-cycle (upstream-downstream) of acceleration. When we also consider the second half-cycle (downstream-upstream) and take the average over the flux, we obtain the expression for the average momentum increase%
	\begin{equation}
		\displaystyle\frac{\Delta p}{p}=\displaystyle\frac{4}{v}\frac{u\rs{p1}-u\rs{p2}}{3}.
		\label{bell2:Deltapavegeneral}
	\end{equation}
	The presented equation accounts for the fact that the plasma velocity profiles are not constant upstream and downstream, contrary to what is assumed in the classical picture of the test-particle acceleration at the shock. 
	This equation shows that the momentum change depends on the difference between the two utmost values of the flow speed that are `reachable' for accelerating particles with a given momentum.
	
	In the Sect.~\ref{bell2:sectNLA}, the probability for a particle to cross the shock from downstream to upstream and to participate consequently in the next acceleration cycle is given by the ratio of the particle fluxes to the positive and to the negative $x$ directions for the isotropic distribution of $\textbf{v}$ and uniform profile of the plasma velocity downstream $u_2(x)=\mathrm{const}$ \citep{1991SSRv...58..259J}. In our case, $u_2$ varies with the distance from the shock. 
	
	Let us consider a simple and intuitive assumption that  particles with a given momentum $p$ can reach the maximum coordinate $x\rs{p2}$ downstream of the shock and this distance is bigger for particles with larger momenta. 
	Particles must return to the shock from this point to continue acceleration. Therefore, a component of a particle velocity $v\rs{x}$ must be larger than the flow speed $u\rs{p2}\approx u_2(x\rs{p})$ there. With these assumptions, the we can write the expression for the probability from (\ref{bell2:probdefEJ}) as
	\begin{equation}
		P=\displaystyle\left(\frac{1-u\rs{p2}/v}{1+u\rs{p2}/v}\right)^2.
		\label{bell2:Pdefnonuni}
	\end{equation}
	The probability for $n$ cycles is
	\begin{equation}
		P=\displaystyle\prod_{i=0}^{n-1}\left(\frac{1-u\rs{p2,\mathit{i}}/v_i}{1+u\rs{p2,\mathit{i}}/v_i}\right)^2.
		\label{Bell2:Probab}
	\end{equation}
	Decomposition of the logarithm to the first order on $u\rs{p2,\mathit{i}}/v_i$ gives  
	\begin{equation}
		\ln P=-\displaystyle\sum_{i=0}^{n-1}\frac{4u\rs{p2,\mathit{i}}}{v_i}
		\label{bell2:Probdecompos}
	\end{equation}
	Appendix~\ref{bell2:Pderiv} presents a more general derivation of this expression.
	
	Let us now substitute this equation with $4/v_i$ from (\ref{bell2:Deltapavegeneral})
	and convert the sum $\sum \Delta p_i/p_i$ to the integral $\int dp'/p'$. As a result, we derive the probability of a particle being accelerated to momentum $p$ (or higher) on a steady background of the nonuniform profiles of the flow velocity upstream and downstream   
	\begin{equation}
		P=\exp\left(-\int_{p\rs{o}}^{p}\frac{3u\rs{p'2}}{u\rs{p'1}-u\rs{p'2}}\frac{dp'}{p'}\right).
		\label{bell2:probability}
	\end{equation}
	
	The isotropic distribution function $f\rs{o}(p)$ is defined by equation (\ref{bell:deffP}) with $n$ the number density of all accelerated particles, making it a fraction $n=\eta n_0$ of all the incoming particles.   
	The flux is conserved $n(x)u(x)=\mathrm{const}$. Therefore, we can take $n$ at any point, e.g. we use $n=\eta n_1 u_1/u\rs{p2}$ to resemble the formula (\ref{bell:fnla}). 
	Thus, the definition (\ref{bell:deffP}) finally yields the distribution function 
	\begin{equation}
		f\rs{o}(p)=\frac{\eta n_1}{4\pi p\rs{o}^3}\frac{3u_1}{u\rs{p1}-u\rs{p2}}
		\exp\left[-\int_{p\rs{o}}^{p}\frac{3u\rs{p'1}}{u\rs{p'1}-u\rs{p'2}}\frac{dp'}{p'}\right]. 
		\label{bell2:fnlayauongu}
	\end{equation}
	
	Known solutions may be easily derived from the equation (\ref{bell2:fnlayauongu}). Namely, $u_2(x)=\mathrm{const}$ results in the non-linear solution of \citet{2002APh....16..429B}. If $u_1(x)=\mathrm{const}$ as well, then we have the common test-particle power-law spectrum for the momentum distribution of accelerated particles. 
	
	For the cosmic-ray modified shock, the way to calculate $u\rs{p1}$ with an integro-differential equation which accounts for the regulation of the flow hydrodynamics by $f\rs{o}(p)$ is developed by \cite{2002APh....16..429B,2005MNRAS.361..907B}. In these references, the flow is assumed to follow the adiabatic law in the pre-shock region. \citet{2006MNRAS.371.1251A,2009MNRAS.395..895C} examine the  impact of other effects in the precursor, such as turbulent heating and magnetic field amplification, on the plasma velocity profile upstream.
	
	The equation (\ref{bell2:fnlayauongu}) describes the momentum distribution of accelerated particles at the shock and is derived in the frame of the approach of \citet{1978MNRAS.182..147B} to the particle acceleration. The solution of the same problem derived in the frame of an alternative approach, namely by solving the kinetic equation for DSA, is presented in Appendix \ref{bell2:kineq}.

	\subsection{Spectral index}
	\label{bell2:sectspind}
	
	The `local' spectral index $s(p)$, i.e. the local slope of the particle distribution $f\rs{o}(p)$ at a given momentum $p$, is defined as
	\begin{equation}
		s(p)=-\frac{d\ln f\rs{o}}{d\ln p}.
	\end{equation}
	Applying it to the function (\ref{bell2:fnlayauongu}), we have
	\begin{equation}
		s(p)=\frac{3u\rs{p1}}{u\rs{p1}-u\rs{p2}}
		+\frac{d\ln(u\rs{p1}-u\rs{p2})}{d\ln p}.
		\label{bell2:spind}
	\end{equation}
	This expression represents the spectral index for particles accelerated around the shock with spatially non-uniform distribution of the flow speed $u(x)$ upstream and downstream. 
	
	\begin{figure}
		\centering 
		\includegraphics[width=0.48\textwidth]{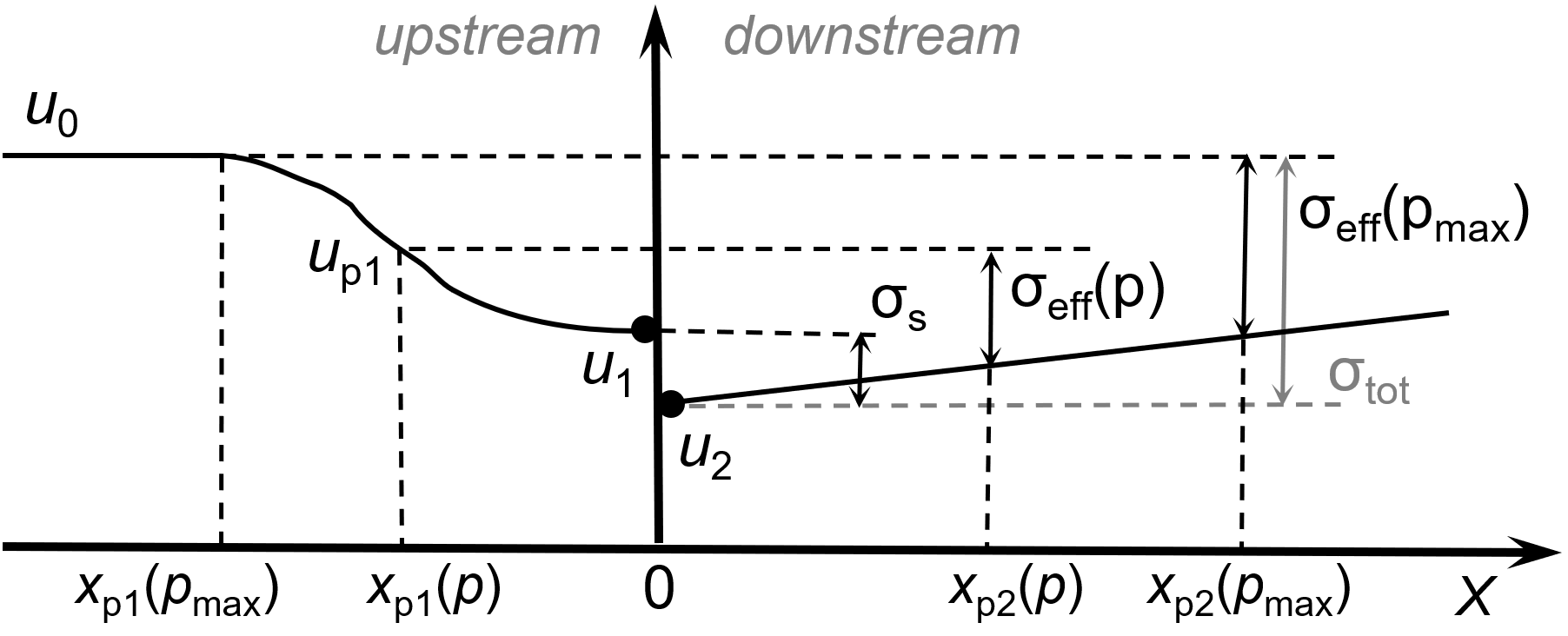}
		\caption{Schematic representation of the structure of the flow velocity $u(x)$ (solid line). The shock is located at the coordinate $x=0$. The upstream profile of $u(x)$ is shown for the case of the cosmic-ray modified shock.  
		}
		\label{bell2:figsxema}
	\end{figure}
	
	Let us introduce the effective compression factor that is `seen' by particles with the momentum $p$ (as shown in Fig.~\ref{bell2:figsxema}):
	\begin{equation}
		\sigma\rs{eff}(p)=\frac{u\rs{p1}}{u\rs{p2}}
		\approx\frac{u_1(x\rs{p1})}{u_2(x\rs{p2})}.
		\label{bell2:sigmaeff}
	\end{equation}
	The spectral index for the cosmic-ray modified shock \cite[it is given by equation (8) in][]{2002APh....16..429B} comes from (\ref{bell2:spind}) by assuming $u_2=\mathrm{const}$ and therefore $\sigma\rs{eff}=u\rs{p1}/u_2$:
	\begin{equation}
		s(p)=\frac{3\sigma\rs{eff}}{\sigma\rs{eff}-1}
		\left(1+\frac{1}{3}\frac{p}{\sigma\rs{eff}}\frac{d\sigma\rs{eff}}{d p}\right).
		\label{bell2:spindBlasi2002}
	\end{equation}
	Equation (\ref{bell2:spind}) gives the classic test-particle index 
	\begin{equation}
		s=\frac{3\sigma\rs{s}}{\sigma\rs{s}-1}
		\label{bell2:calssind}
	\end{equation}
	for $u_1=\mathrm{const}$ and $u_2=\mathrm{const}$, namely, by substitution $\sigma\rs{eff}=\sigma\rs{s}=u_1/u_2$. 
	In the general case when $u_1(x)\neq\mathrm{const}$ and $u_2(x)\neq\mathrm{const}$, the contribution to $s(p)$ from the first term in (\ref{bell2:spind}) is the same as for the unmodified shock (\ref{bell2:calssind}) but with the effective compression factor $\sigma\rs{eff}$ instead of $\sigma\rs{s}$. This contribution is modified somehow by the second term in (\ref{bell2:spind}), which represents the `local' (in the momentum space) slope of the difference $u\rs{p1}-u\rs{p2}$. 
	
	Let us consider the effects of the spatially nonuniform $u_2(x)$ on a few practical cases. 
	
	\subsection{Benchmark problems}
	\label{bell2:sectpract}
	
	\subsubsection{Test particles at Sedov shock}
	
	We start from the acceleration of the test particles, i.e.   $u_1=\textrm{const}$, at adiabatic shock with the Sedov profile for $u_2(x)$. 
	
	For simplicity and clarity, we use the approximation by \citet{1950RSPSA.201..159T} for the distribution of the flow velocity downstream of the \citet{1959sdmm.book.....S} shock. It is 
	\begin{equation}
		v(r)=v_2(R) r/R 
	\end{equation}  
	in the laboratory frame, the immediate post-shock value is $v_2(R)=3V/4$, $V$ is the shock speed, $r$ the distance from the explosion center, $R$ the shock radius. 
	The accuracy of the approximation is better than $4\%$ for $>0.7R$ \citep[figure~1 in][]{2000A&A...357..686P}. In the shock reference  frame, the profile is
	\begin{equation}
		u_2(x)=u_2(0)\cdot (1+3x/R)
		\label{bell2:sedovu2}
	\end{equation}
	where $u_2(0)=u_1/\sigma\rs{s}$. 
	We calculate $u\rs{p2}$ with expression $u\rs{p2}\approx u_2(x\rs{p})$ which  follows from the definition (\ref{bell2:defu2p}) and ${\cal X}_2(x)\approx {\cal H}(x\rs{p}-x)$. In order to parametrise the dependence of the diffusion length  $x\rs{p}$ on $p$, we take 
	\begin{equation}
		x\rs{p}(p)\propto\frac{D_2(p)}{u_2(0)}
	\end{equation}
	where $D$ is the spatially constant diffusion coefficient of the form $D_2\propto p^\alpha$ with $\alpha\geq 0$. Therefore, 
	\begin{equation}
		u\rs{p2}=u_2(0)\cdot(1+3\bar x\rs{p}),\qquad
		\bar x\rs{p}(p)=\bar x\rs{pm}\cdot\left({p}/{p\rs{max}}\right)^\alpha 
		\label{bell2:equp2xp}
	\end{equation}
	where $\bar x\rs{pm}\equiv x\rs{p}(p\rs{max})/R$ is the diffusion length for particles with the maximum momentum $p\rs{max}$ normalized to $R$.
	In such a model, we can evaluate the spectral index analytically:
	\begin{equation}
		s(p)=\frac{3\sigma\rs{s}-3\alpha\bar x\rs{p}(p)}{\sigma\rs{s}-1-3\bar x\rs{p}(p)}.
		\label{bell2:spindsedov}
	\end{equation}
	
	\begin{figure}
		\centering 
		\includegraphics[width=\columnwidth]{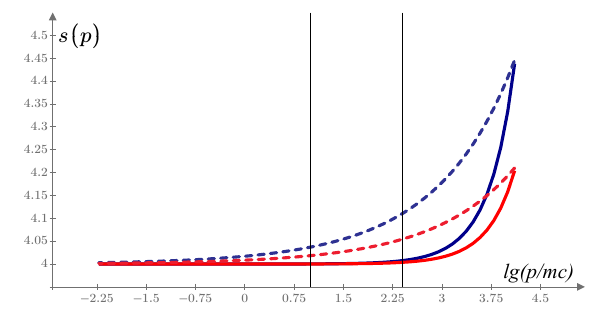}\\
		\includegraphics[width=\columnwidth]{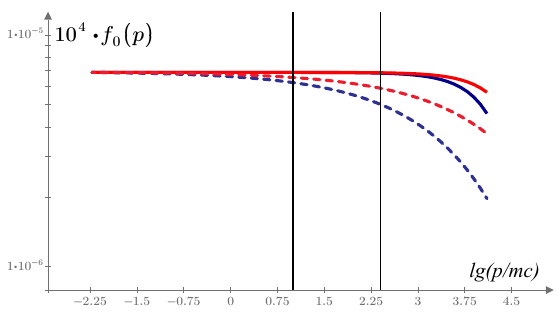}
		\caption{The local slopes $s(p)$ and the distribution functions $f\rs{o}(p)/n\rs{o}$ for the unmodified Sedov shock. 
			The parameters are $\bar x\rs{pm}=0.1$ (blue lines), $\bar x\rs{pm}=0.05$ (red lines), $\alpha=1$ (solid lines), $\alpha=1/3$ (dashed lines), $p\rs{max}=10^4m\rs{p}c$. The classic value (for $u_2=\mathrm{const}$) is $s=4$ and is independent of $p$, $\alpha$ and $\bar x\rs{pm}$. 
			Vertical lines: left -- the energy of electrons for which the radio index is calculated in Fig.~\ref{bell2:figsevolindex}; right -- protons with energy $240\un{GeV}$ (see Sect.~\ref{bell2:HESBr}).
		}
		\label{bell2:figscase1}
	\end{figure}

	To see the sensitivity of the problem to $\bar x\rs{pm}$, we consider a few values of this parameter. We take these values for granted in the present section and explore their role further in Sect.~\ref{bell2:sect-xpdiscuss}. 
	
	In Fig.~\ref{bell2:figscase1}, we can see the local spectral index and the distribution function for the model we are considering. It is evident that the effect of the function $u_2(x)$, which increases with $x$, leads to a higher  $s$ for the higher $p$. (Note that in the observer frame, the flow speed $v$ decreases downstream with distance from the shock.) It is becase the effective compression factor $\sigma\rs{eff}$ decreases with growing $p$ (Fig.~\ref{bell2:figsxema}). Therefore, the momentum distribution function $f\rs{o}(p)$ descends more rapidly than a common power-law (Fig.~\ref{bell2:figscase1}). 
	
	It's worth noting that the shapes of the particle distributions shown in Fig.~\ref{bell2:figscase1} are only due to the plasma velocity gradient downstream of the shock. We applied only the sharp high energy cutoff to the spectrum of particles ${\cal H}(p\rs{max}-p)$, where ${\cal H}(x)$ is the Heaviside step function. For comparison, the spectrum is just the power law up to the maximum momentum $N(p)=Kp^{-s}{\cal H}(p\rs{max}-p)$ in case of the constant flow speed $u_2(x)=\mathrm{const}$.
	
	The classical test-particle index for the case of $u_2=\mathrm{const}$ is $s=4$. From Fig.~\ref{bell2:figscase1}, we observe that the local index $s(p)$ for the Sedov velocity profile is close to this value for particles with low momenta $p$, i.e. for the particles which are accelerated from the small distances $\bar x\rs{p}\approx 0$, where deviations of the flow velocity from $u_2(0)$ are minor. The index $s(p)$ increases with $p$ for Sedov shock, and its maximum value happens for particles around $p\rs{max}$ which are able to return to the shock from deeper downstream locations, i.e. from around $\bar x\rs{pm}$.
	
	The effect described in the present paper is sensitive to $\alpha$. For a fixed $\bar x\rs{pm}$, the effect is more prominent for smaller $\alpha$ (Fig.~\ref{bell2:figscase1}) because particles with low momenta may return to the shock from deeper $\bar x\rs{p}$ downstream regions. 
	
	Electrons in the magnetic field $B=10\un{\mu G}$ emitting most of their energy through the synchrotron radiation at the radio frequency $5\un{GHz}$ have energy $10\un{GeV}$. It is apparent from Fig.~\ref{bell2:figscase1} that, at this energy, the spectral index differs by a few percent from the usual value. For example, the radio index is $0.52$ for $\alpha=1/3$ and $\bar x\rs{pm}=0.1$ instead of classic $0.5$. The differences are larger for smaller $\alpha$. 
	In the limit of a momentum-independent $D$ (i.e. $\alpha=0$), the spectral index is also momentum independent. In such a toy model, its value depends on the distance $x\rs{p}$ which is the same for any $p$, as it follows from the equation (\ref{bell2:equp2xp}). Numerically, for the Sedov profile of $u_2(x)$, the values are $s=4.4$ for $\bar x\rs{pm}=0.1$ and $s=4.2$ for $\bar x\rs{pm}=0.05$. The radio index at $5\un{GHz}$ is $0.7$ and $0.6$ respectively.  
	
	\subsubsection{Cosmic-ray modified Sedov shock}
	
	Let us consider the acceleration of particles on the cosmic-ray modified shock with the upstream profile $u_1(x)\neq\mathrm{const}$ and the downstream distribution for $u_2(x)$ given by equation (\ref{bell2:sedovu2}), i.e. like in a Sedov SNR. The goal is to compare the spectra of particles with the case $u_2(x)=\mathrm{const}$. 
	
	The essence of the non-linearity of the cosmic-ray modification of the shock is in the dependence of the plasma speed profile upstream of the shock $u_1(x)$ on the particle distribution function $f\rs{o}(p)$. If we have $u_2(x)=\mathrm{const}$ downstream of the shock, then the NLA solution of \cite{2002APh....16..429B} holds. In our problem, we should account for the spatial variation $u_2(x)$ as well. We show above that the non-uniform profile $u_2(x)$ affects the shape of the distribution function $f\rs{o}(p)$. Therefore, the profile  $u_1(x)$  upstream will be modified by cosmic rays differently compared to the uniform $u_2(x)$ distribution. In turn, this modified upstream profile affects $f\rs{o}(p)$, too.
	
	\begin{figure}
		\centering 
		\includegraphics[width=\columnwidth]{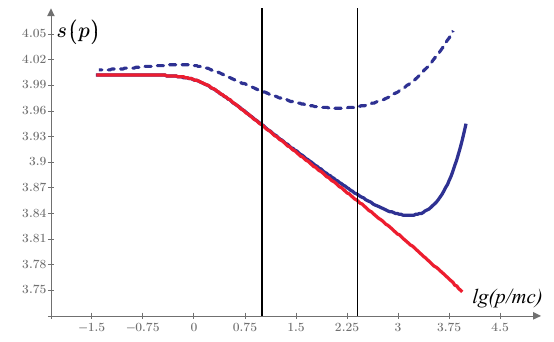}\\
		\includegraphics[width=\columnwidth]{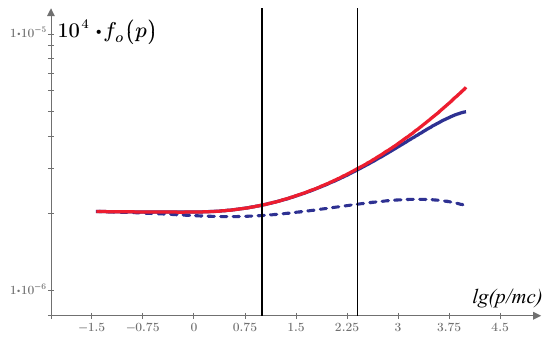}
		\caption{Plots of $s(p)$ and $f\rs{o}(p)/n\rs{o}$ for the cosmic-ray modified shock with $u_2(x)=\mathrm{const}$ (red line) and with $u\rs{p2}(p)$ given by the equation (\ref{bell2:equp2xp}) with
			$\alpha=1$ (solid blue line), $\alpha=1/3$ (dashed blue line), $\bar x\rs{pm}=0.1$, $p\rs{max}=10^4m\rs{p}c$.
		}
		\label{bell2:figscase2}
	\end{figure}
	
	We calculated the NLA solution for such a problem numerically in a self-consistent way by following the approach described by \citet{2002APh....16..429B,2005MNRAS.361..907B}. 
		Namely, we found the function $u\rs{p1}(p)$ by solving iteratively the integro-differential equation which depends on $f\rs{o}(p)$, accounts for the adiabatic evolution of the flow in the shock precursor, for the relation between the injection efficiency and the compression factor at the sub-shock and -- our addition -- for the flow velocity structure downstream. Then the distribution function $f\rs{o}(p)$ may be found from the equation (\ref{bell2:fnlayauongu}). The iterations are repeated until the boundary condition $u\rs{p1}(p\rs{max})=u_0$ is satisfied. 
	We considered the following set of parameters: the Mach number ${\cal M}\rs{o}=100$, the flow speed far upstream $u_0=6.5\E{8}\un{cm/s}$, the maximum momentum $p\rs{max}=10^4m\rs{p}c$ and the injection efficiency $\eta=1.30\E{-4}$. The red line on Fig.~\ref{bell2:figscase2} shows the solution for  $u_2(x)=\mathrm{const}$ while the blue lines correspond to  models with the velocity gradient downstream of the shock.
	
	Qualitatively, we can understand the behavior of the momentum distribution of accelerated particles from   Fig.~\ref{bell2:figsxema}. In this scheme, we show a model of the cosmic-ray modified shock (it is represented by the velocity profile in the $x<0$ domain) with increasing function $u_2(x)$ downstream (in the $x>0$ domain). If we first consider the uniform $u_2(x)$ (dashed grey horizontal line on Fig.~\ref{bell2:figsxema}), we can observe that particles with the lowest momenta `see' the compression $\sigma\rs{s}$. Particles with larger momenta are affected by continuously higher compression $\sigma\rs{eff}(p)$ which raises due to the velocity profile upstream. It is the main characteristic of the NLA solution shown by the red line in Fig.~\ref{bell2:figscase2}: the higher the effective compression, the lower the spectral index $s$ and the harder the distribution $f\rs{o}(p)$. However, the increasing function $u_2(x)$ (solid line downstream in Fig.~\ref{bell2:figsxema}) progressively weakens this trend for higher momenta. As a result, the spectrum's shape may be less concave compared to the pure NLA case or have a complex shape with an inflection point (blue lines in Fig.~\ref{bell2:figscase2}). In contrast, the decreasing $u_2(x)$ causes stronger effective compression compared to the uniform $u_2$ downstream and strengthens the NLA concavity. 
	
	The effective compression factors of are typically high in the NLA problem. They often lead to the steep shapes of the emission spectra which are not observed in SNRs. To make $\sigma\rs{eff}$ lower compared to the pure NLA case,  \citet{2009MNRAS.395..895C} consider the relative Alf\'venic speed $v\rs{A}$ of the scattering centres (i.e. $u+v\rs{A}$ instead of $u$ in the diffusion-convection equation). Our model with a rising flow speed in the shock reference frame could be a natural way to lower $\sigma\rs{eff}$ and make the spectrum flatter in circumstances where the cosmic-ray shock modification is an essential component. 
	
	It's worth noting that the injection efficiency we consider for Fig.~\ref{bell2:figscase2} is not high, and the shock modification is not very strong. Indeed, the compression factor $\sigma\rs{tot}\equiv u_0/u_2(0)$ is $5.434$ for the red line, $5.359$ for the blue solid line and $4.897$ for the blue dashed line. If the back-reaction of accelerated particles is higher, e.g. $\sigma\rs{tot}$ is about $10$ or more, then the role of the upstream profile $u_1(x)$ in the shaping $f\rs{o}(p)$ dominates the effect from the downstream distribution $u_2(x)$.

	\section{Discussion}
	\label{bells:dicsus}
	
	We saw in the previous section that the spatial variation of the flow speed downstream of the shock results in deviation of the momentum distribution of accelerated particles from the classic power law. The spectrum of emission produced by these particles should also be affected.  
	Could we see such effect in some observables? How strong could it be, if any? Let us look for an answer in the frame of the test-particle acceleration, i.e. to assume in the present section that $u_1=\textrm{const}$. 
	
	\subsection{Evolution of the radio index of young SNRs}
	
	Previously, we explored magneto-hydrodynamic properties of young SNRs on examples of remnants from the three types of supernovae \citep{2021MNRAS.505..755P}. Sect.~\ref{bell2:sectHD} shows (see also an animation in Appendix~\ref{bell2:appmovie}) that the temporal changes in the plasma velocity structure downstream of the shock $u_2(x,t)$ are essential during the development of SNR from the early stage \citep{1982ApJ...258..790C,1998ApJ...497..807D} to the \citet{1959sdmm.book.....S} epoch. These changes may cause time variations in the momentum distribution of particles and, consequently, in the evolution of the radio index of SNRs.
	
\begin{figure}
	\centering 
	\includegraphics[width=\columnwidth]{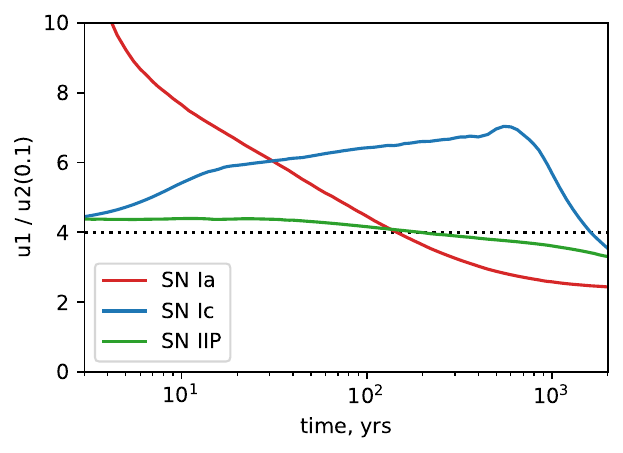}
	\caption{The effective compression factor $u_1/u_2(\bar x\rs{pm})$ with $\bar x\rs{pm}=0.1$ for different ages. Models of SNRs are the same as in Sect.~\ref{bell2:sectHD}. 
	}
	\label{bell2:figsevolsigmaeff}
\end{figure}

	Fig.~\ref{bell2:figsevolsigmaeff} shows the evolution of the effective compression factor for particles with the maximum momentum $\sigma\rs{eff}(p\rs{max})=u_1/u_2(\bar x\rs{pm})$ with $\bar x\rs{pm}=0.1$ for three models of SNRs. The flow dynamics are different between the models. Specifically, the flow speed within $\bar x\rs{pm}$ from the shock is almost the same in the SNIIP model, which is one of the representative cases for the DSA theory that assumes $u_2(x)=\mathrm{const}$. On the other hand, the SNIa model demonstrates prominent changes in conditions for the shock particle acceleration during its evolution. 
	
	These differences may be present in the evolution of the radio index. To expose the effect, we approximate 
	the profile of $u_2(x,t)$ for the downstream region close to the shock by the linear function with a variable coefficient $k$:
	\begin{equation}
		u_2(x,t)=u_2(0)\cdot \left(1+k(t)x/R\right).
		\label{bell2:sedovuyoung}
	\end{equation}
	This approximation is justifiable by the shape of the $u_2(x)$ profiles within the interval of $x$ from zero to $\sim 0.1R$ (Fig.~\ref{bell2:figshdv} and animation in Appendix~\ref{bell2:appmovie}).
	Then the function $u\rs{p2}(p)$ is
	\begin{equation}
		u\rs{p2}(p)=u_2(0)\cdot\left(1+k(t)\bar x\rs{p}(p)\right)
		\label{bell2:equp2xp2}
	\end{equation}
	instead of (\ref{bell2:equp2xp}) and 
	the local slope of the particle distribution is, instead of (\ref{bell2:spindsedov}),
	\begin{equation}
		s(p,t)=\frac{3\sigma\rs{s}-k(t)\alpha\bar x\rs{p}(p)}{\sigma\rs{s}-1-k(t)\bar x\rs{p}(p)}.
		\label{bell2:spindyoung}
	\end{equation}

	\begin{figure}
		\centering 
		\includegraphics[width=\columnwidth]{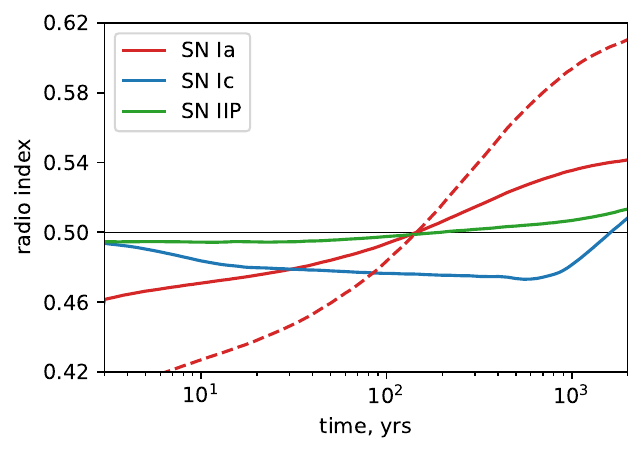}
		\caption{The evolution of the radio index at $5\un{GHz}$ for the three models of SNRs and electrons with energy $10\un{GeV}$ (so, the magnetic field strength is $10\un{\mu G}$), $p\rs{max}=10^4m\rs{p}c$, $\bar x\rs{pm}=0.1$, $\alpha=1/3$ (solid lines). The dashed line represents the same model as the solid line but with artificially taken $\alpha=1/5$. 
		}
		\label{bell2:figsevolindex}
	\end{figure}
	
	The evolution of the radio index $(s-3)/2$ calculated with this expression for $s(p,t)$ is shown in Fig.~\ref{bell2:figsevolindex}. 
	In the model SNIa displaying the most prominent changes in the plasma speed's structure (Fig.~\ref{bell2:figsevolsigmaeff}), the radio index varies from  $0.46$ to $0.54$ during about $1000\un{yrs}$  (red solid line, $s$ changes from $1.9$ to $2.1$ respectively). The range for the radio index variation could be even wider if particles probe deeper interiors of SNR during acceleration (i.e. when ($x\rs{pm}>0.1R$ or $\alpha$ smaller in the assumed dependence $x\rs{p}(p)$, Eq.~\ref{bell2:equp2xp}), as shown by the dashed line in Fig.~\ref{bell2:figsevolindex}. 
	It is also consistent with Fig.~\ref{bell2:figscase1}. 
	
	We can conclude that the spectral index in the SNIa model is smaller than the value $0.5$ up to SNR age $\approx 100\un{yrs}$ and higher at later times, as shown in Fig.~\ref{bell2:figsevolindex}. The reason is given by 
	Equation~(\ref{bell2:spindyoung}), which shows that the classic value $s=4$ (and the radio index $0.5$) corresponds to $k=0$. Furthermore, this formula reveals that $s$ could be below $4$ when $k<0$ (which happens when the flow velocity $u_2$ decreases with $x$, and thus the effective compression factor $u_1/u_2(\bar x\rs{pm})>4$, as seen in Fig.~\ref{bell2:figsevolsigmaeff}) or above $4$ when $k>0$.\footnote{This is valid for $\alpha<3\sigma\rs{s}/(\sigma\rs{s}-1)$ that is always the case.} 
	
	It's worth noting that the left vertical line in Fig.~\ref{bell2:figscase1} corresponds to electrons that mostly determine the radio index shown in Fig.~\ref{bell2:figsevolindex}. We see that the deviation of their spectrum from the power law is minor. However, the deviation is more significant for particles with higher momenta that emit X-rays and gamma-rays.

	\subsection{High-energy spectral break}
	\label{bell2:HESBr}
	
	A spectral feature is observed in the gamma-ray spectra of several SNRs. It is attributed to a change in the slope of the proton spectrum which is typically referred to as the broken power law.  
	The change in the slope $\Delta s\approx 0.4-1$ and should be added to proton spectrum models around the break momentum $p\rs{br}$, which has values from $p\rs{br}=10\un{GeV}/c$ to $p\rs{br}=240\un{GeV}/c$ to fit the observed spectra of some SNRs \citep[e.g.][]{2013Sci...339..807A,2015A&A...574A.100H,2018A&A...612A...5H}. 
	Such a feature is also visible in radio \citep{2019MNRAS.482.3857L} and microwave \citep{2016A&A...586A.134P} spectra of SNRs, although it has roots in the momentum distribution of electrons. 
	
	Our results suggest that the particles accelerated at the shock with spatially variable flow velocity downstream have a momentum distribution with a smooth change of slope over a wide range of momenta. 
	The radio-to-X ray synchrotron spectrum is convex for evolved SNRs (when $u_2(x)$ is increasing function, Sect.~\ref{bell2:sectHD}) that could mimic the `break' if it is not very sharp. 
	Our calculations in Sect.~\ref{bell2:sectspind} show that
	the slope changes between the particle energies $1\un{Gev}$ and $300\un{Gev}$ as $\Delta s=0.11$ (from $s=4.02$) or $\Delta s=0.30$ (from $s=4.3$) for Sedov shock and parameters $(\bar x\rs{pm},\alpha)$ from $(0.1,0.3)$ to $(0.2,0.1)$. Comparing these values of $\Delta s$ with those required by observations, we see that the model could (at least partially) be responsible for the `broken power law'. 
	
	The difference of $s(p)$ from the standard value $s=4$ increases toward the maximum energy, faster for larger $\bar x\rs{pm}$ (i.e. how far from the shock the particles can return back), for the higher value of $k$ (i.e. how fast the flow speed $u_2$ grows with $x$) and for the lower $\alpha$ (i.e. how the diffusion coefficient depends on the particle momentum). 
	
	The last point leads to an important conclusion: the shape of the accelerated particle distribution in the momentum space is sensitive to the dependence of $D$ on $p$ if the flow speed is not spatially uniform. It is clear from equation (\ref{bell2:spindyoung}): the larger the absolute value of $k$, the higher the sensitivity. 
	
	\begin{figure}
		\centering 
		\includegraphics[width=\columnwidth]{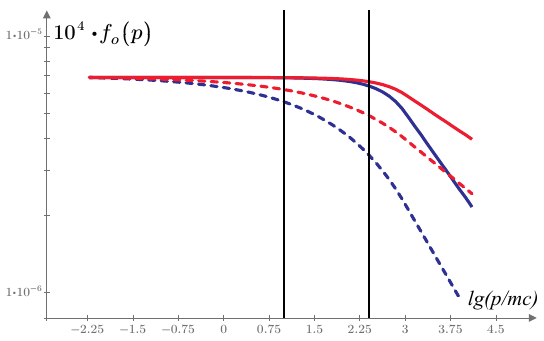}
		\caption{Distribution functions for the same cases as in Fig.~\ref{bell2:figscase1}.
			The only difference is the diffusion coefficient. It is the same as in Fig.~\ref{bell2:figscase1} up to the momentum $0.1p\rs{max}$ and independent of $p$ for higher momenta. Above $0.1p\rs{max}$, the slopes $s$ for the solid blue and red lines are $2.33$ and $2.16$ respectively. 
			Note that all the lines will be straight horizontal up to $p\rs{max}$ in case of the uniform flow speed $u_2(x)=\mathrm{const}$.
		}
		\label{bell2:figdifcoef}
	\end{figure}
	
	In our model, the diffusion coefficient is a power-law function of the particle momentum, $D\propto p^{\alpha}$ across the whole range of $p$. As a result, the momentum spectrum of particles changes smoothly over the wide range of $p$. The observed spectral feature may reflect the change of dependence of the diffusion coefficient on $p$ around the break momentum $p\rs{br}$. 
	
	Fig.~\ref{bell2:figdifcoef} illustrates this concept, showing the same models as Fig.~\ref{bell2:figscase1} with the only difference in the diffusion coefficient for particles with the highest momenta. Namely, it is 
	\begin{equation}
		D(p)\propto \left\{
		\begin{array}{ll}
			p^{\alpha},& p<p\rs{br}\\
			p\rs{br}^{\alpha}=\mathrm{const},& p\geq p\rs{br}
		\end{array}
		\right.
	\end{equation}
	with $p\rs{br}=0.1p\rs{max}$. We see a sharp break around $p\sim10^3/mc$ in the solid lines representing the Bohm diffusion up to the break. The slope change is $\Delta s=0.33$ for $x\rs{pm}=0.1$ and $\Delta s=0.16$ for $x\rs{pm}=0.05$, meaning that it depends on the depth of the `zone' downstream which is sampled by the accelerating particles. It is interesting to note that the slope $s=2.33$ above $p\rs{br}$ around the lower boundary of the values of $s\approx 2.3-2.6$ is required to fit the observed gamma-ray spectra of some SNRs at the photon energies $\gtrsim 1\un{TeV}$. The index $s$ may even be higher in our model for appropriate $x\rs{pm}$ and $k$. 
	
	We would like to emphasize that $D(p)$ influences $s(p)$ only if the flow speed is non-uniform, in either downstream or upstream direction. If $u_2(x)=\mathrm{const}$ and $u_1(x)=\mathrm{const}$, then particles of all momenta sample the same difference between $u_1$ and $u_2$ and, therefore, $s$ is independent of $p$.

		\subsection{Evolution of the diffusion length}
		\label{bell2:sect-xpdiscuss}
		
		The effect we are considering depends on the spatial extent of the downstream region probed by particles accelerating at the shock. We expect that the effect from the non-uniform flow velocity downstream of the shock may be noticeable in real SNRs if the diffusion length $x\rs{pm}$ is higher than say $5\%$ of the radius $R$.
		
		The Larmor radius of a proton or electron with energy $E=E\rs{TeV}\cdot 10^{12}\un{eV}$ in the magnetic field $B=B\rs{\mu}\cdot 10^{-6}\un{G}$ is $r\rs{L}=10^{-3}E\rs{TeV}/B\rs{\mu}\un{pc}$. The length-scale downstream involved into formation of $f\rs{o}(p)$ is of order of the diffusion length $x\rs{p}\simeq D_2/u_2$. If diffusion is Bohm-like, then $D\rs{B}=\eta r\rs{L} c/3$ and
		\begin{equation}
			x\rs{p}\simeq \frac{4\eta c}{3V} r\rs{L}.
			\label{bell2:xpDugen}
		\end{equation} 
		The diffusion length could be up to $x\rs{p}\sim 10^{5} r\rs{L}$ (for $\eta=100$ and a slow shock $V=300\un{km/s}$). Under such extreme circumstances, even the radio-emitting electrons may effectively feel the non-uniformity of the flow speed distribution because the diffusion length may be up to $x\rs{p}\sim 0.1 \un{pc}$ for $10\un{GeV}$ electrons in $10\un{\mu G}$ magnetic field. 
		Particles with energy $E\rs{max}=10\un{TeV}$ may have $x\rs{p}\sim 0.1\un{pc}$ in more common situations, such as in a magnetic field with $50\un{\mu G}$ strength behind the shock with speed $6000\un{km/s}$ and $\eta=10$. 
		
		In order to see the sensitivity of the problem to the parameter $\bar x\rs{pm}=x\rs{p}(p\rs{max})/R$, we adopt its fixed values for calculations in the previous sections. Let us examine how $\bar x\rs{pm}$ evolves in the two alternative scenarios: the efficient acceleration with the Bohm  diffusion and the test-particle one with the Kolmogorov diffusion coefficient. 
		
		The turbulent component of the magnetic field $\delta B$ causes the scattering of particles. The equation (\ref{bell2:xpDugen}) yields
		\begin{equation}
			x\rs{p,B}\simeq \frac{\eta\sigma\rs{tot} c}{3\sigma\rs{B}Ve\delta B_1}E  = \frac{\eta\sigma\rs{tot}}{30\sigma\rs{B}} \frac{E\rs{TeV}}{V\rs{3000}\delta B\rs{\mu}}\ \un{pc}
			\label{bell2:xpDu}
		\end{equation} 
		where $V\rs{3000}$ is the shock speed in units $3000\un{km/s}$ and we used $\delta B_2=\sigma\rs{B}\delta B_1$. It follows that the normalized diffusion length varies over time due to the expansion of SNR, the evolution of $p\rs{max}$, and the magnetic field: $\bar x\rs{pm}\propto p\rs{max}/(VBR)$. We know the dependencies $R(t)$ and $V(t)$ for our SNR models from simulations. 
		
\begin{figure}
			\centering 
			\includegraphics[width=\columnwidth]{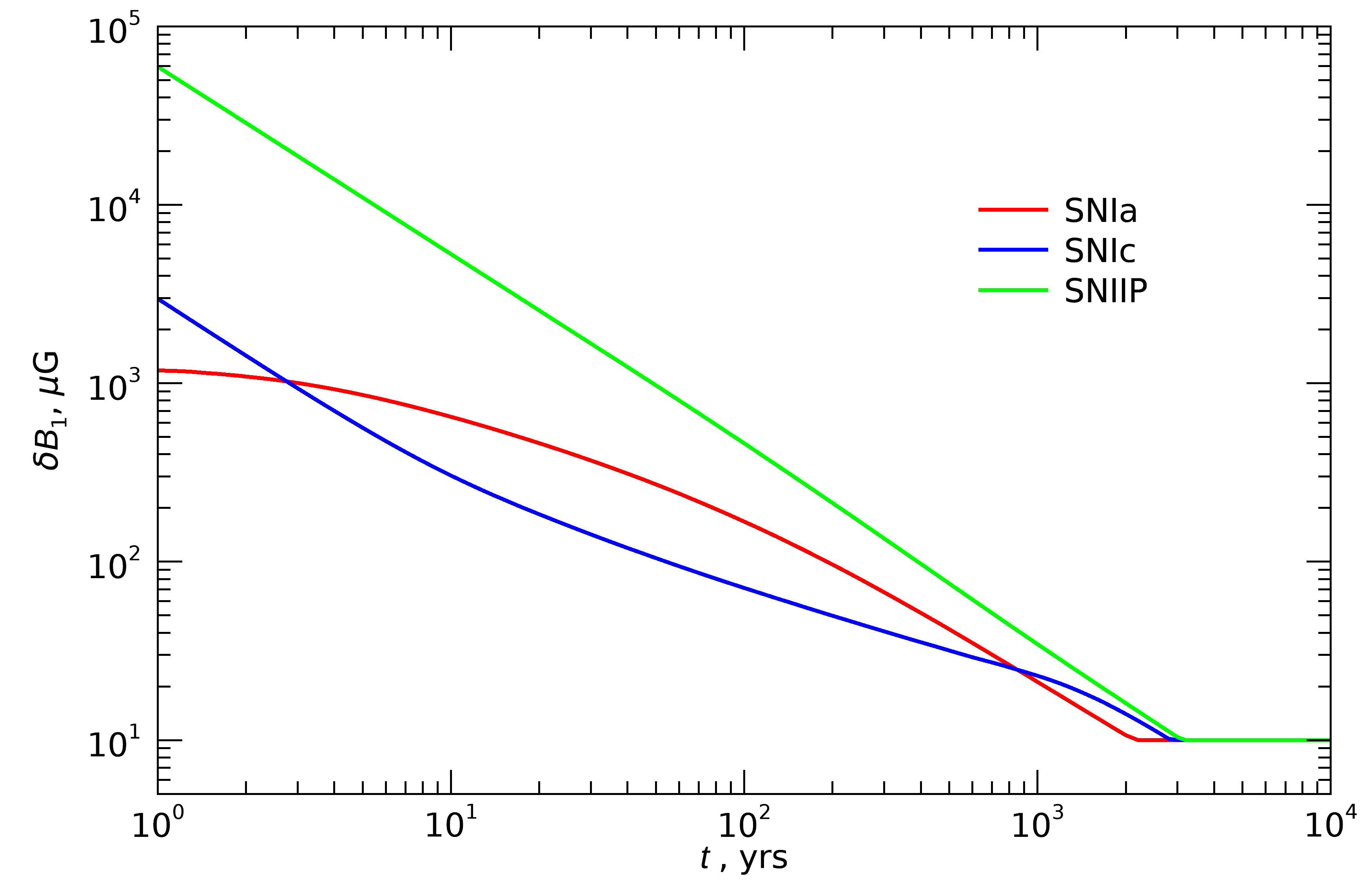}
			\caption{Evolution of $\delta B_1$ in three models of SNRs.
			}
			\label{bell2:figdBB}
		\end{figure}

		Under the efficient acceleration regime, we consider both the ordered $B_1$ and the random $\delta B_1$ components. At the young shocks, the Bell non-resonant instability operates with $\delta B_1>B_1$. The saturated level of magnetic energy is about the $V/c$ fraction of energy $E\rs{cr}$ stored in CRs \citep{2004MNRAS.353..550B}
		\begin{equation}
			\frac{\delta B_1^2}{4\pi}\simeq\frac{V}{c}\xi\rs{cr}\rho\rs{o}V^2
			\label{bell2:deltaBsat}
		\end{equation}
		where $\xi\rs{cr}$ is the fraction of the kinetic energy transferred to CRs. 
		When this $\delta B_1$ is less than the strength of the ordered field $B_1$, the resonant waves are responsible for the particle scatterings with \citep{2006MNRAS.371.1251A} 
		\begin{equation}
			\frac{\delta B_1^2}{8\pi}\simeq \frac{v\rs{A}}{V}P\rs{cr}
			\label{bell2:deltaBres}
		\end{equation}
		where $v\rs{A}=B_1/\sqrt{4\pi\rho\rs{o}}$ and $P\rs{cr}=(\gamma\rs{cr}-1)E\rs{cr}$  the CR pressure, $\gamma\rs{cr}=4/3$ and $\delta B_1/B_1\leq 1$. 
		\op{The highest value of $\delta B_1$ for the resonant waves could be $B_1$. We assume $\delta B_1=B_1$ in situations where these formulae give, at the same time, simultaneously for the nonresonant $\delta B_1<B_1$ and the resonant $\delta B_1>B_1$.}
		Numerically, for \op{the plasma with the mean mass of particle in units of the proton mass} $\mu=0.609$,
		\begin{equation}
			\delta B_1= \left\{
			\begin{array}{ll}
				110\ \xi\rs{cr}^{1/2} n\rs{o}^{1/2}V\rs{3000}^{3/2}\ \un{\mu G},& \delta B_1>B_1\\[8pt]
				27\ \xi\rs{cr}^{1/2} B\rs{\mu}^{1/2}n\rs{o}^{1/4}V\rs{3000}^{1/2}\ \un{\mu G},& \delta B_1< B_1\\[8pt]
				B_1 & \mathrm{otherwise}
			\end{array}
			\right.
			\label{bell2:deltaBnum}	
		\end{equation}
		Fig.~\ref{bell2:figdBB} shows the variation of $\delta B_1$ with the SNR age in our models. 
		
		The mean acceleration time $t\rs{acc}(p\rs{max})$ could be estimated from a classic integral \citep[eq.~3.31 in][]{1983RPPh...46..973D} assuming $D\propto p^\alpha$, $u_1=V$, $u_2=V/\sigma\rs{tot}$:
		\begin{equation}
			t\rs{acc}(p\rs{max})\simeq\frac{\beta\rs{a}}{\alpha}\frac{D(p\rs{max})}{V^2},
			\quad\beta\rs{a}=\frac{3\sigma\rs{tot}(1+\sigma\rs{tot}/\sigma\rs{B})}{\sigma\rs{tot}-1}
			\label{bell2:tacc}
		\end{equation}
		where $\beta\rs{a}$ accounts for the fractions of times a particle spends upstream and downstream.
		As expected, the acceleration time is longer in $1/\alpha$ times compared to the case of the Bohm diffusion ($\alpha=1$), e.g. in 3 times for the Kolmogorov diffusion ($\alpha=1/3$).
		
		One of three processes determines the maximum momentum of accelerated particles $p\rs{max}$. 
		The condition $t\rs{acc}(p\rs{max})=t$ gives the maximim momentum limited by the SNR age. We have for the Bohm-like diffusion
		\begin{equation}
			p\rs{max,age,B}=\frac{3e}{\eta\beta\rs{a}}\left(\frac{V}{c}\right)^2\delta B_1t.
			\label{bell2:pmaxage}
		\end{equation}
		The synchrotron loss time, applicable for electrons only, is
		\begin{equation}
			t\rs{rad}=\frac{1}{A\rs{p}\langle B^2 \rangle p},\quad 
			A\rs{p}=\frac{4}{3}\frac{\sigma\rs{T}}{8\pi}\frac{1}{(m\rs{e}c)^2}
			=4.7\E{7}\un{cgs}
		\end{equation}
		with $\langle B^2 \rangle=\beta\rs{b} B\rs{rad}^2$ where $B\rs{rad}=\max(\delta B_1,B_1)$ and $\beta\rs{b}$ accounts for the times particles spend in the upstream and downstream \citep{1998ApJ...493..375R}
		\begin{equation}
			\beta\rs{b}=\frac{\sigma\rs{B}+\sigma\rs{B}^2\sigma\rs{s}}{\sigma\rs{B}+\sigma\rs{s}}.
		\end{equation}
		The equation $t\rs{acc}(p\rs{max})=t\rs{rad}(p\rs{max})$ provides 
		\begin{equation}
			p\rs{max,rad,B}=\left[\frac{3e}{\eta\beta\rs{a}\beta\rs{b}A\rs{p}}\left(\frac{V}{c}\right)^2\frac{\delta B_1}{B\rs{rad}^2}\right]^{1/2}.
			\label{bell2:pmaxrad}
		\end{equation}
		The third important time to account is the time-to-grow for the Bell instability that produces the magnetic field \citep{2013MNRAS.431..415B,2015APh....69....1C}. 
		In this case, the maximum momentum is estimated by the equation~(5) in \citet{2020APh...12302492C}:
		\begin{equation}
			p\rs{max,gr}\simeq\frac{\xi\rs{cr}e\pi^{1/2}}{5\Lambda} 
			\frac{4}{4-\omega}
			R \rho\rs{o}^{1/2}\left(\frac{V}{c}\right)^2
			\label{bell2:pmaxgr}
		\end{equation}
		where $e$ is the electron charge, $\omega=-d\ln n\rs{o}(R)/d\ln R$ is the ambient density steepness \citep[it is denoted by $m$ in][]{2015APh....69....1C}, $\Lambda\approx 12$ for $p\rs{max}\sim 100\un{TeV}$. This expression is valid for any type of diffusion and when the non-resonant $\delta B_1>B_1$. 
		
		\begin{table}
			\centering
				\caption{We consider two scenarios for particle acceleration to calculate the evolution of $p\rs{max}$ and $\bar x\rs{pm}$. The type of particles is different to show the role of the radiative losses in one case and to remove it from analysis in another scenario for simplicity.}
				\label{tab:bell2:table}
				\begin{tabular}{c|cc} 
					\hline
					& efficient & test-particle \\
					& acceleration & regime \\
					\hline
					particle & proton & electron \\		
					diffusion & Bohm & Kolmogorov \\		
					$\alpha$ & $1$ & $1/3$ \\		
					$\eta$ & $1$ & $10$ \\
					$\xi\rs{cr}$ & 0.15 & -- \\
					$E\rs{o}$ & -- & $80\un{GeV}$ \\
					$\delta B_1$ & (\ref{bell2:deltaBnum}) & $B_1$ \\
					$B\rs{rad}$ & $\max(\delta B_1,B_1)$ & $B_1$ \\
					$B_1$ & $10\un{\mu G}$ & $10\un{\mu G}$ \\
					$\sigma\rs{B}$ & $\sqrt{11}$ & 1 \\
					$\sigma\rs{tot}$ & 5.5 & 4 \\
					$p\rs{max}$ & min (\ref{bell2:pmaxage}), (\ref{bell2:pmaxgr}) & min (\ref{bell2:pmaxageKolm}), (\ref{bell2:pmaxradKolm})\\
					$\beta\rs{a}$ & 9.7 & 20 \\	
					$\beta\rs{b}$ & 7.2 & 1 \\
					\hline
			\end{tabular} 
			\label{bell2:table2scen}
		\end{table}
		\begin{figure}
			\centering 
			\includegraphics[width=\columnwidth]{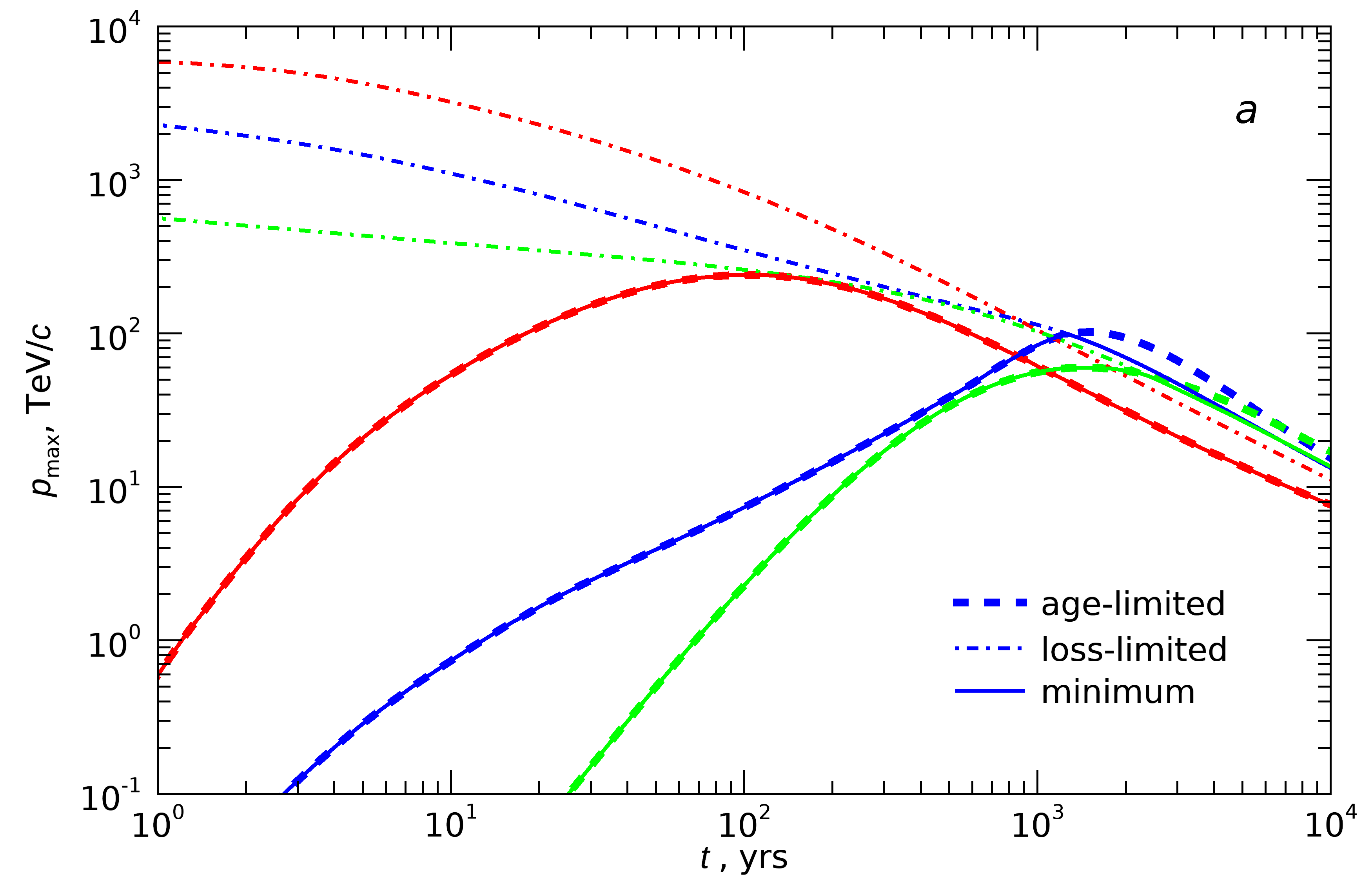}\\
			\includegraphics[width=\columnwidth]{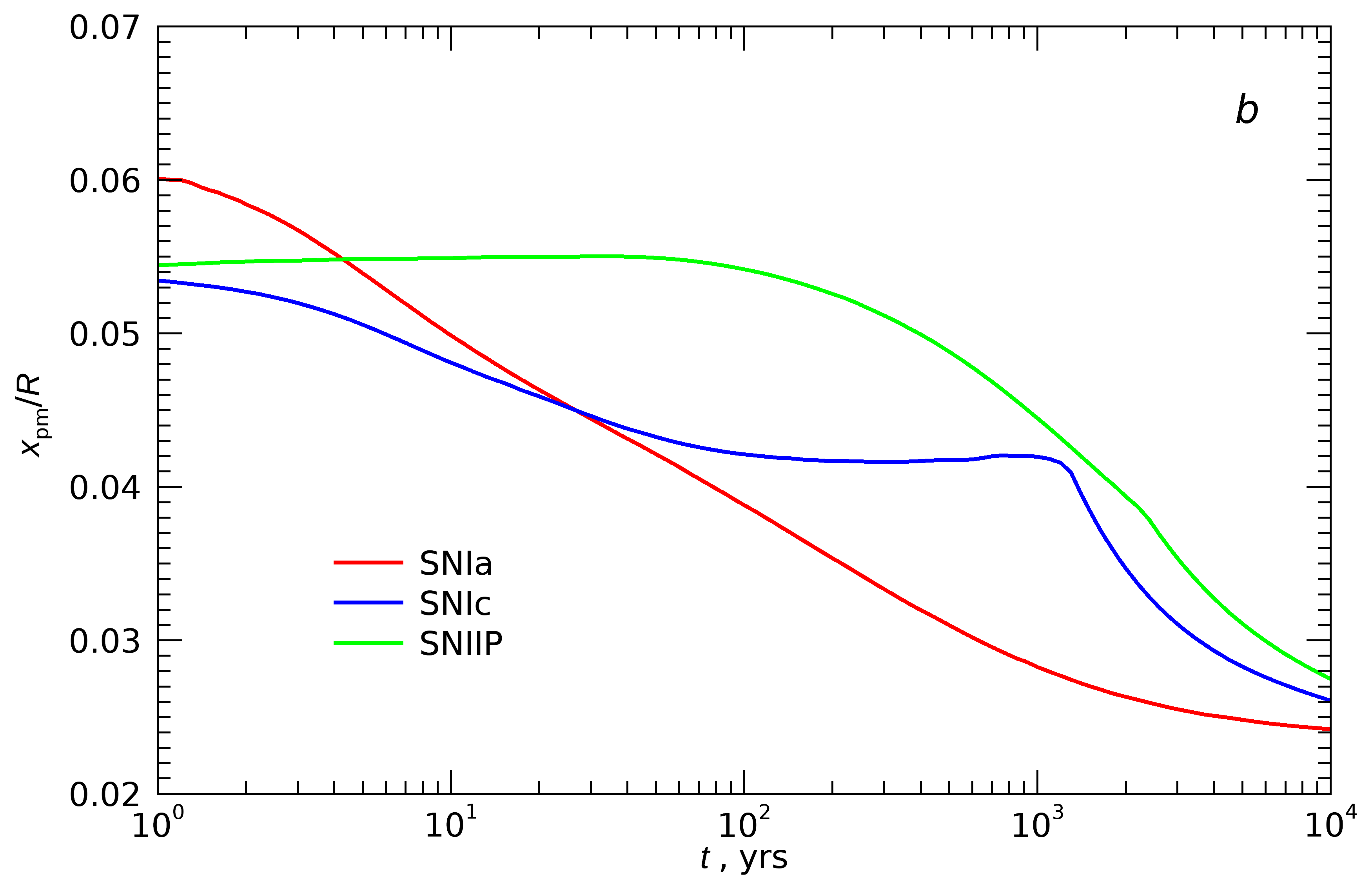}
			\caption{Evolution of the maximum momentum $p\rs{max}$ \textbf{(a)} and the normalized diffusion length $\bar x\rs{pm}$ \textbf{(b)} for particles with the maximum momentum in the test-particle acceleration scenario with Kolmogorov diffusion. Different colours show three of our models of SNRs. Line types represent the time-limited (dashed lines), loss-limited (dot-dashed lines) $p\rs{max}$ and the minimum of the two (solid lines). Lines for $\bar x\rs{pm}$ correspond to the solid lines on the plot for the maximum momenta.  $p\rs{max,rad}<p\rs{max,age}$ from $t=1.3\un{kyr}$ for the blue and $t=2.3\un{kyr}$ for the green solid lines. 
			}
			\label{bell2:figpmaxTP}
		\end{figure}
		\begin{figure}
			\centering 
			\includegraphics[width=\columnwidth]{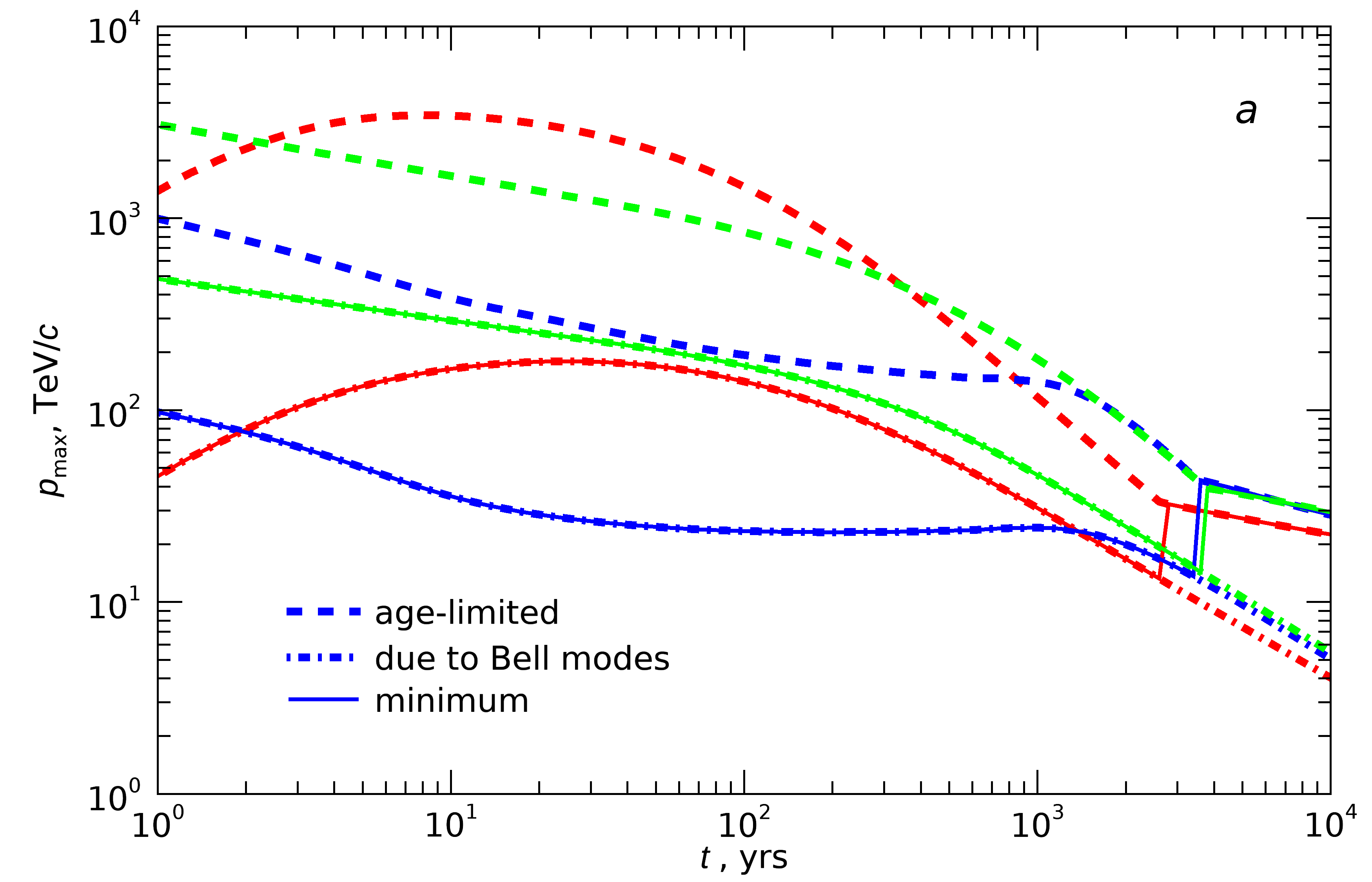}\\
			\includegraphics[width=\columnwidth]{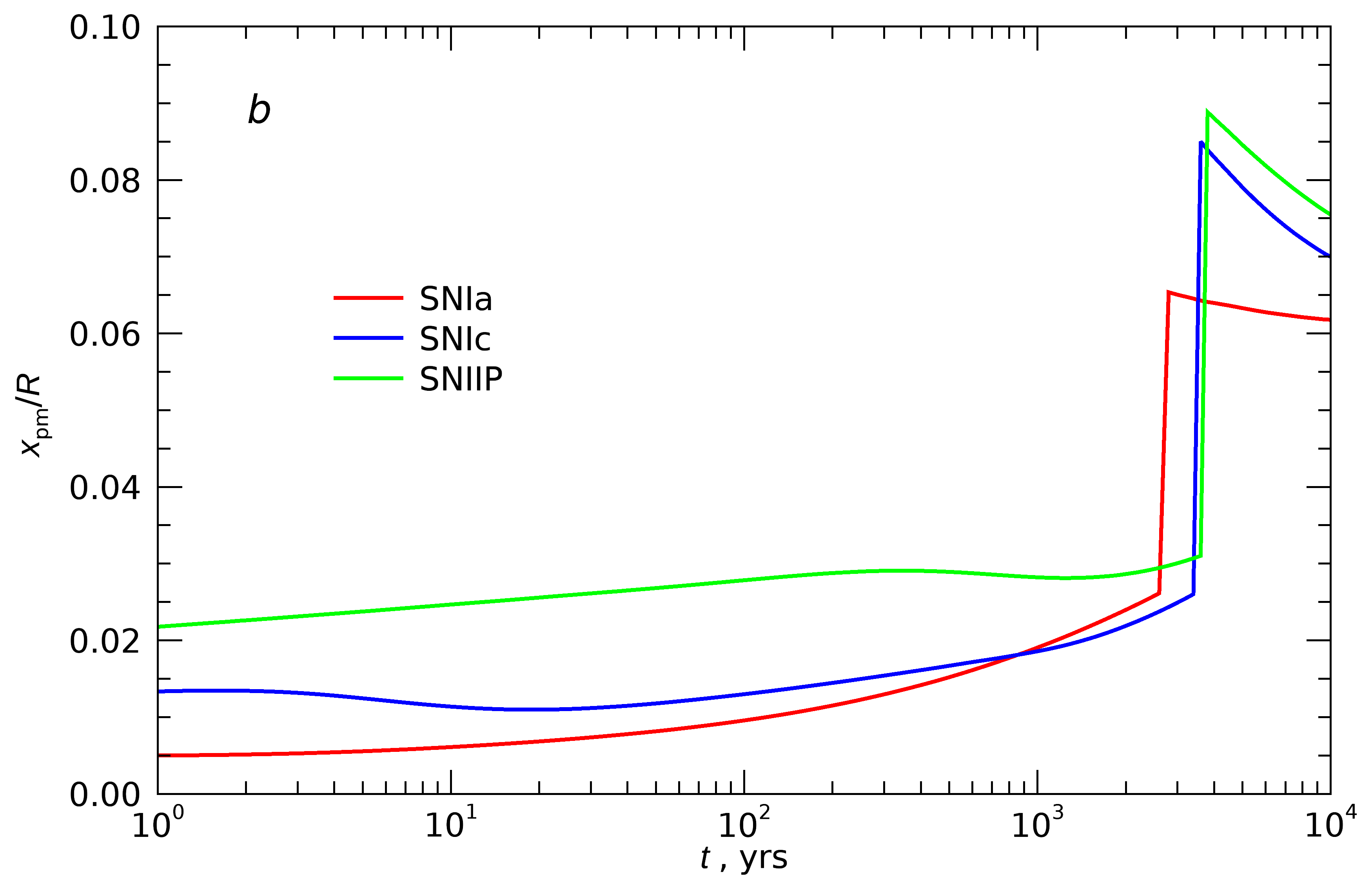}
			\caption{The same as in Fig.~\ref{bell2:figpmaxTP} for the scenario of efficient acceleration. Dot-dashed lines represent $p\rs{max}$ due to the Bell instability which operates up to the time when the non-resonant $\delta B_1>B_1$.
			}
			\label{bell2:figpmaxNL}
		\end{figure}

		In an alternative scenario, we assume the test-particle acceleration and the Kolmogorov diffusion. 
		We set the diffusion coefficient the same as the Bohm-like one at the particle energy $E\rs{o}$, i.e. $D\rs{K}=\eta c E\rs{o}^{2/3}E^{1/3}/(3e\delta B)$. Then 
		\begin{equation}
			x\rs{p,K}\simeq \frac{\eta\sigma\rs{tot} c}{3\sigma\rs{B}Ve\delta B_1} E\rs{o}^{2/3}E^{1/3} = 
			\frac{\eta\sigma\rs{tot}}{30\sigma\rs{B}} \frac{ E\rs{o,TeV}^{2/3}E\rs{TeV}^{1/3}}{V\rs{3000}\delta B\rs{\mu}}\ \un{pc}
			\label{bell2:xpDuK}
		\end{equation} 
		\begin{equation}
			p\rs{max,age,K}=\left[\frac{ec^{2/3}}{\eta \beta\rs{a} E\rs{o}^{2/3}}\left(\frac{V}{c}\right)^2B_1t\right]^{3}.
			\label{bell2:pmaxageKolm}
		\end{equation}
		\begin{equation}
			p\rs{max,rad,K}=\left[\frac{ec^{2/3}}{\eta\beta\rs{a}\beta\rs{b}A\rs{p}  E\rs{o}^{2/3}}\left(\frac{V}{c}\right)^2 \frac{\delta B_1}{B\rs{rad}^2}\right]^{3/4}.
			\label{bell2:pmaxradKolm}
		\end{equation}
		`Test-particle' relativistic protons do not amplify the magnetic field. Therefore, we do not account for $p\rs{max,gr}$ in this scenario. 
		MF strength for the diffusion coefficient and the radiative losses is the ordered one at the parallel shock $\delta B_1=B\rs{rad}=B_1=B_2$. 
		
		The two scenarios for the evolution of the diffusion length $\bar x\rs{pm}=x\rs{pm}/R$ are summarized in Table~\ref{bell2:table2scen}. Figs.~\ref{bell2:figpmaxTP}, \ref{bell2:figpmaxNL} show results of calculations.
		
		In the test-particle case, the maximum momentum in our models is limited predominantly by the age of the system (Fig.~\ref{bell2:figpmaxTP}a). Interestingly, the three lines for $\bar x\rs{pm}$ (Fig.~\ref{bell2:figpmaxTP}b) have shapes similar to the evolution of the deceleration parameter $m=Vt/R$ \citep[figure 2 in ][]{2021MNRAS.505..755P}. Indeed, if we express $D$ from the relations $t\rs{acc}=t$ or $t\rs{acc}=t\rs{rad}$ and substitute $\bar x\rs{pm}=\sigma\rs{tot}D(p\rs{max})/(RV)$ with it, we get 
		\begin{equation}
			\bar x\rs{pm}=\left\{
			\begin{array}{ll}
				\displaystyle\frac{\alpha\sigma\rs{tot}}{\beta\rs{a}}m(t), & t\geq t\rs{rad}\\[8pt]
				\displaystyle\frac{\alpha\sigma\rs{tot}}{\beta\rs{a}}\frac{t\rs{rad}(p\rs{max})}{t}m(t), & t< t\rs{rad}\\
			\end{array}
			\right.
			\label{bell2:xpmnorm}
		\end{equation} 
		that is valid for diffusion of any type. If  the age of the systems limits $p\rs{max}$, then the correlation $\bar x\rs{pm}(t)\propto m(t)$ is clear. When the radiative losses limit the maximum energy of particles, then $\bar x\rs{pm}$ scales also with parameters determining $t\rs{rad}$ (Fig.~\ref{bell2:figpmaxTP}b, blue line after $t=1.3\un{kyr}$, green line after $t=2.3\un{kyr}$). It could be essential in real SNRs when radiative losses limit $p\rs{max}$, e.g. when MF is amplified to high values. Then  $\bar x\rs{pm}\propto B^{-1/2}t^{8/5}$ for a Sedov SNR as an example, that might provide a way to determine properties of magnetic field from the non-thermal spectra of SNRs. 
		When $p\rs{max}$ is age-limited, the normalized diffusion length for particles with maximum momentum is determined solely by the shock compression and diffusion type (represented by $\alpha$). In particular, for the Bohm-like (Kolmogorov) diffusion and unmodified shock, $\bar x\rs{pm}\approx 20\ (6)\%$ of the SNR radius in the young SNRs (when $m\approx 1$) or $\bar x\rs{pm}\approx 8\ (3)\%$ at the Sedov stage (when $m=0.4$). 
		
		In the scenario of efficient acceleration, $p\rs{max}$ is limited by the grow-rate of the Bell instability up to $t\approx 3\un{kyr}$ (Fig.~\ref{bell2:figpmaxNL}a, cf. Fig.~\ref{bell2:figdBB}). Eqs.~(\ref{bell2:xpDu}), (\ref{bell2:deltaBsat}) and (\ref{bell2:pmaxgr}) yield then 
		\begin{equation}
			\bar x\rs{pm}=0.025\ \frac{\eta\sigma\rs{tot}}{\sigma\rs{B}}\frac{4}{4-\omega}
			\frac{\xi\rs{cr}^{1/2}}{V\rs{3000}^{1/2}}\sim 0.02\eta V\rs{3000}^{-1/2}.
		\end{equation}
		The dependence on the shock speed is not strong, and the evolution of $\bar x\rs{pm}$ is slow. Indeed, we see in Fig.~\ref{bell2:figpmaxNL}b that the region downstream involved into the acceleration process is about $1\div 3\%$ of the radius when the Bell instability is effective. However, if the diffusion is Bohm-like with $\eta=2\div3$, then $\bar x\rs{pm}$ could reach $5\%$ of the SNR radius. At times after $\sim 3\un{kyr}$, when the non-resonant instability is not effective in scattering the particles, then the SNR age limits the maximum momentum. At this stage, the diffusion length does not depend on $\eta$ (Eqs.~\ref{bell2:deltaBres} and \ref{bell2:pmaxage}) and may be up to $\sim 0.1R$.
		
		Thus, the downstream structure of the flow might have some influence on the shape of the non-thermal spectra of SNRs in both scenarios we are considering. 
		The effect is not strong enough but may modify the spectral indices in realistic cases under some circumstances.

	\subsection{Limitations}
	\label{bell2:sect-limit}
	
	The non-uniformity of the plasma speed behind the shock affects the cosmic-ray spectrum to a larger extent if particles can penetrate deeply downstream while accelerating at the shock. If so, they can probe a larger (for decreasing $u_2(x)$) difference between the upstream $u\rs{p1}$ and downstream $u\rs{p2}$ flow speeds and get a higher bust in momentum $\Delta p\propto u\rs{p1}-u\rs{p2}$. The distance from the shock depends on the properties of diffusion, $x\rs{p2}\simeq D/u_2$. Therefore, in order the effect from the downstream gradient of $u_2$ be visible in the spectrum, the diffusion coefficient should be large. In our estimates we considered $x\rs{p1}$ to be $5\div 10\%$ of the SNR radius $R$ for the particles with the maximum momentum $p\rs{max}=10^4m\rs{p}c\approx  10\un{TeV}/c$. This corresponds to a diffusion coefficient for particles with such momentum $D\sim 0.1R u_2\sim 3\E{25}\un{cm^2/s}$ (we used $R\sim1\un{pc}$ and $u_2\sim10^3\un{km/s}$ for this estimate), that is well within typical values used in the literature. 
	On the contrary, to be capable of providing the proton acceleration to energies beyond $100\un{TeV}$, the diffusion coefficient should be the smallest one, namely, the Bohm one $D=r\rs{L}c/3$ in the magnetic field of few $100\un{\mu G}$, that requires substantial magnetic field amplification. Under such conditions, the diffusion length for particles with energy $E\rs{max}=10\un{TeV}$ is $D_2/u_2\simeq r\rs{L}c/u_2\sim 0.01\un{pc}$ that is just $\sim 1\%$ of $R$ for particles with the maximum momentum. Of course, the variation of the plasma speed provided by the SNR hydrodynamics on such length scales downstream does not affect the shape of the CR spectrum. On the other hand, the length-scale $x\rs{p}\propto D\propto p\rs{max}^\alpha$ also increases with rise of the maximum momentum. 
	
	In general, we expect that the effect of the downstream velocity gradient could be prominent in SNRs or their regions where the particle acceleration is not very efficient, i.e. with a lower magnetic field and higher diffusion coefficient. Thus, in the present paper, we are not in the conditions to provide the highest $E\rs{max}\sim \un{PeV}$ energies. Instead, the maximum energy $E\rs{max}\approx 10\un{TeV}$ we used in our examples is a value which may be reached in SNRs with a common interstellar magnetic field and without the fastest diffusion regime. It seems that the conditions for the very efficient acceleration (amplified $B\sim\mathrm{few}\un{mG}$, the smallest Bohm diffusion coefficient), while crucial for the investigation of PeVatrons, are not widespread in SNRs. Such conditions could be expected locally, i.e. present only in some small regions of the SNR shock surface. As illustration of this point, we refer the reader to the figure~4 in \citet{2003ApJ...586.1162L}. The thinnest radial profile of the X-ray surface brightness in region A of SN1006 is due to the very fast radiative cooling of electrons in the high magnetic field. This profile provides evidence of effective acceleration with magnetic field amplification there. However, two other profiles B and C, being even nearby, are thicker (i.e. magnetic field is lower in these regions). Therefore, the high magnetic strength inferred from the thinnest profile may not be generally attributed to the whole SNR. The less efficient particle acceleration could be in operation throughout other regions of the SNR shock surface. 
	
	Another point is related to the dependence of the diffusion coefficient on the particle momentum, i.e. the value of the index $\alpha$. The effect from the gradient of $u_2$ could be more prominent at the smaller particle momenta if the index $\alpha$ is less than unity (like in the Kolmogorov or Kraichnan diffusion coefficients) because $x\rs{p}(p)\propto x\rs{p}(p\rs{max}) p^{\alpha}$ (see Figs.~\ref{bell2:figscase1}, \ref{bell2:figscase2}, \ref{bell2:figdifcoef}).
	
	Interestingly, numerical studies of the non-linear particle acceleration (where the kinetic particle equation was solved, along with the system of the hydrodynamic equations for the flow) by different groups \citep{1995ApJ...447..944K,2007APh....28..232K,2000A&A...364..911G,2002ApJ...579..337K,1997APh.....7..183B,2000A&A...357..283B,
		2010ApJ...708..965Z,2012APh....39...12Z} used either the Bohm or the Bohm-like diffusion coefficient and did not find an appreciable impact of the downstream velocity gradient on the particle spectrum.
	
	We found one report where the effect appears visible (though it is difficult to separate from the eventual influence of other factors). \citet{2020A&A...634A..59B} numerically solved a system of equations  for cosmic rays, magnetic turbulence and plasma hydrodynamics in SNRs. Figure 2 in their calculations compares the particle spectra for the Bohm diffusion (red lines) and the diffusion coefficient calculated self-consistently from the magnetic turbulence (green lines). In the latter case, the coefficient becomes Kolmogorov-like for most energies (Brose 2023, private communication). The differences between the spectra shapes for $\alpha=1$ and $\alpha=1/3$ at the ages $10$ and $60\un{kyr}$ are similar to ours in Fig.~\ref{bell2:figscase1}.  
	
	In summary, the effect of the non-uniform $u_2(x)$ on the particle spectrum could be prominent in SNRs where adoption of Kolmogorov ($\alpha=1/3$) or Kraichnan diffusion ($\alpha=1/2$) in magnetic fields with standard strength $B$ of few $10\un{\mu G}$ is acceptable, or, in case of a more efficient acceleration with the Bohm-like diffusion coefficient with $\eta>1$. 
	In contrast, such values of $\alpha$, $B$ and $\eta$ are not compatible with the highest maximum energies of cosmic rays observed in several SNRs and are not consistent with a modern view of anisotropic cascade of MHD turbulence in the sources of highly efficient particle acceleration based on the Bohm diffusion and high values of amplified $B$.
	

	\section{Conclusions}
	\label{bell2:conclusions}
	
	Our research aims to study the possible effect of the spatially variable distribution of the flow speed downstream of the shock on the momentum distribution of particles accelerated at the shock. 
	
	To achieve this, we generalized the approach of \citet{1978MNRAS.182..443B} which is an alternative way to describe the particle acceleration at the shock. This result is of general academic interest of its own.
	
	We applied this generalized individual particle approach to the NLA solution of \citet{2002APh....16..429B}, which assumes $u_2(x)=\mathrm{const}$ and then to a more general problem with $u_2(x)\neq\mathrm{const}$. 
	
	In order to cross-check our results, we derived in Appendix~\ref{bell2:kineq} the same general solution for $f\rs{o}(p)$ by solving the kinetic equation. 
	
	The formula for the general solution (\ref{bell2:fnlayauongu}) is similar to the known NLA solution. The difference consists in substitution $u\rs{p2}(p)$ instead of the constant $u_2$. 
	
	In Sect.~\ref{bell2:sectHD}, we present profiles of the flow speed derived in numerical MHD simulations of remnants evolving after supernova explosions of different types. We demonstrate that the flow speed may vary prominently on the spatial scales involved in the acceleration of particles around the SNR shocks. 
	
	Then, we calculate the momentum spectra of particles for several practical cases to demonstrate the effect and its limitations. 
	In these applications, we treat the problem by taking the formula (\ref{bell2:equp2xp2}) for the function $u\rs{p2}(p)$. 
	%
	Such a simplified description of $u\rs{p2}(p)$ allows us to study the effect of a non-uniformity of the flow speed on the particle spectrum accelerated in the test-particle limits or the NLA regime (Sect.~\ref{bell2:sectpract}). 
	
	
	In Sect.~\ref{bells:dicsus}, we suggest two cases where the phenomenon might affect, at least partially, some observed properties of SNRs. 
	
	The overall effect may be visible from the equation (\ref{bell2:spindyoung}) written for the test-particle limit. If the flow speed in the frame of the shock rises with distance from the shock (i.e. $k>0$ like in evolved SNRs) and particles with higher $p$ reach larger $x$, then $s$ grows with $p$. This results to a convex spectrum of particles. If the flow speed declines ($k<0$), then $s$ decreases and the spectrum is concave. The typical assumption of the uniform flow downstream  ($k=0$) results in the classic formulae. 
	
	An important conclusion apparent from the equation (\ref{bell2:spindyoung}) is that the diffusion coefficient (represented by $\alpha$ in this formula) directly affects the shape of $f\rs{o}(p)$ only if the flow speed is not uniform upstream and/or downstream. 
	Diffusion properties of accelerating particles are important in shaping the $f\rs{o}(p)$ because the diffusion coefficient determines the diffusive lengths for particles with different momenta. The momentum increase of a particle is sensitive to the difference of the flow velocities it probes $u\rs{1}(x\rs{p1})-u\rs{2}(x\rs{p2})$. Therefore, the same speed profile $u_2(x)$ affects the spectrum $f\rs{p}(p)$ in a different way in models with distinct diffusion coefficients. 
	
	The effect of interest is rather sensitive to the diffusion properties of cosmic rays. Particles should diffuse deeply downstream to probe a hydrodynamic variation of $u_2(x)$. Therefore, their diffusion length and diffusion coefficient should be high.  
	As the acceleration of cosmic rays in SNRs to energies beyond $100\un{TeV}$ is likely achievable only through the Bohm diffusion mechanism in an amplified magnetic field (then the diffusion length is small), the significance of the downstream gradient of plasma velocity in SNRs, capable of generating PeV cosmic rays, may generally be disregarded.

	\begin{acknowledgements}
		We acknowledge Elena Amato for useful discussions. 
		OP acknowledges the OAPa grant number D.D.75/2022 funded by Direzione Scientifica of Istituto Nazionale di Astrofisica, Italy. The study was partially supported by the INAF mini-grant 1.05.23.04.04. This project has received funding through the MSCA4Ukraine project, which is funded by the European Union. Views and opinions expressed are however those of the author(s) only and do not necessarily reflect those of the European Union. Neither the European Union nor the MSCA4Ukraine Consortium as a whole nor any individual member institutions of the MSCA4Ukraine Consortium can be held responsible for them. We thank the Armed Forces of Ukraine for providing security to perform this work. This research used an HPC cluster at DESY, Germany, and the HPC system MEUSA at INAF-Osservatorio Astronomico di Palermo, Italy.
	\end{acknowledgements}
	
	\bibliographystyle{aa}
	\bibliography{bell_young}

	
	
	\begin{appendix}  
		
		\section{Evolution of the flow velocity downstream of the forward shock. Animation}
		\label{bell2:appmovie}
		
		The animation shows the time development of the 
		profiles of the flow speed $v(r)$ in the observer reference frame (left) and $u(x)$ in the shock reference frame (right) for the same models of SNRs as described in Sect.~\ref{bell2:sectHD} and shown in Fig.~\ref{bell2:figshdv}, up to $10\,000\un{yrs}$. 
		The horizontal axes represent $r$ on the right plot and $x=1-r/R$ on the left plot. 
		Vertical axes are normalized on $v\rs{2}(0)$ (left) and $u_1(0)$ (right).
		The dashed line on each plot corresponds to the flow speed in the \citet{1959sdmm.book.....S} solution.

		\section{Probability of return}
		\label{bell2:Pderiv}
		
		The probability of return $P$ in one acceleration cycle is given by the ratio $P={F_{-x}}/{F_{+x}}$ where $F_{-x}$ and $F_{+x}$ are the fluxes of particles with momentum $p$ in the negative and positive $x$-directions \citep{1991SSRv...58..259J}. In our case, the flow speed varies with  $x$ and we need to integrate over all places where the probability ${\cal X}_2(p,x)$ to find particles with this momentum is above zero, i.e.
		\begin{equation}
			\begin{array}{ll}
				F_{-x}&=\displaystyle\int\limits_{0}^{1}d{\cal X}_2\left|\int\limits_{-v}^{-u_2(x)}dv_x(u_2(x)+v_x)\right| 
				\\&\displaystyle=\frac{v^2}{2}\int\limits_{+\infty}^{+0}
				\left(1-\frac{u_2(x)}{v}\right)^2
				\frac{d{\cal X}_2}{dx}dx\\
				F_{+x}&=\displaystyle\int\limits_{0}^{1}d{\cal X}_2\int\limits_{-u_2(x)}^{v}dv_x(u_2(x)+v_x)
				\\&\displaystyle=\frac{v^2}{2}\int\limits_{+\infty}^{+0}
				\left(1+\frac{u_2(x)}{v}\right)^2
				\frac{d{\cal X}_2}{dx}dx.
			\end{array}
			\label{bell2:appfluxesdefs}
		\end{equation}
		Transition to the rightmost equalities assumes that ${\cal X}_2=1$ at $x=+0$ and ${\cal X}_2=0$ for $x=+\infty$.
		The simplified treatment of the probability in Sect.~\ref{bell2:distrfunc} may be recovered by taking approximately 
		\begin{equation}
			{\cal X}_2(x)={\cal H}(x\rs{p}-x)
			\label{bell2:apprXHeav}
		\end{equation}
		and considering that $d{\cal H}(x\rs{p}-x)/dx=-\delta(x-x\rs{p})$. Then 
		\begin{equation}
			P=\displaystyle\left(\frac{1-u\rs{p2}/v}{1+u\rs{p2}/v}\right)^2
			\label{bell2:Pdefnonuni_appendix}
		\end{equation}
		where $u\rs{p2}=u_2(x\rs{p})$ as it follows from the approximation (\ref{bell2:apprXHeav}) and the definition
		\begin{equation}
			u\rs{p2}\equiv \int_{0}^{1}u_2(x)d{\cal X}_2
			=u_2(0)-\int_{+\infty}^{+0} {\cal X}_2(x) \frac{du_2}{dx}dx.
			\label{bell2:defup2append}
		\end{equation}
		
		In a more general treatment, we neglect the terms of the order $(u_2/v)^2$ in (\ref{bell2:appfluxesdefs}). Then, $(1\pm u_2/v)^2\approx 1\pm 2u_2/v$ and, with the use of the definition (\ref{bell2:defup2append}), the probability of return in one cycle becomes 
		\begin{equation}
			P=
			\frac{1-{2u\rs{p2}}/{v}}
			{1+{2u\rs{p2}}/{v}}.
		\end{equation}
		In many cycles, we have the product of the probabilities for individual cycles. 
		Decomposition of the logarithm of the product into the series to the first order in $u\rs{p2}/v$ yields the equation (\ref{bell2:Probdecompos}).
		
		\section{Solution of the kinetic equation}
		\label{bell2:kineq}
		
		This Appendix presents an alternative approach to derive the expression (\ref{bell2:fnlayauongu}) from the kinetic equation.
		
		We consider the standard steady-state equation for the particle distribution function $f(x,p)$:
		\begin{equation}
			u\pd{f}{x}=\pd{}{x}\left[D\pd{f}{x}\right]+\frac{1}{3}\frac{du}{dx}p\pd{f}{p}+Q\rs{o}(p)\delta(x)
			\label{kinu:maineq}
		\end{equation}
		together with the typical conditions: the flow speed is along the $x$ axis, in the positive $x$ direction; the shock is located at the coordinate $x=0$; the velocity jump $du/dx=(u_2-u_1)\delta(x)$ through the point $x=0$; $f(-0,p)=f(+0,p)$; $f(x,p)=0$ at $x=\pm\infty$ and $df/dx=0$ at $x=-\infty$; the momentum part of the injection term
		\begin{equation}
			Q\rs{o}(p)=\displaystyle\frac{\eta n_1u_1}{4\pi p^2}\delta(p-p\rs{o})
		\end{equation}
		where $p\rs{o}$ is the injection momentum,$\eta$ is the injection efficiency (a fraction of particles which start acceleration), $p\rs{o}$ the injection momentum, $\delta(x)$ is the delta-function, 
		the indices `1' and `2' mark 
		upstream and downstream locations. 
		
		We consider the general situation when the flow speed profiles are not spatially constant: $u_1(x)\neq\mathrm{const}$, $u_2(x)\neq\mathrm{const}$.  
		The approach to solving the equation (\ref{kinu:maineq}) is similar to solving the non-linear problem \citep{2002APh....16..429B,2005MNRAS.361..907B}, i.e. we have to account for the spatial variations of the flow velocity. 
		
		The integration of the main equation (\ref{kinu:maineq}) from $-\infty$ to $x=-0$ results in
		\begin{equation}
			\left[D\pd{f}{x}\right]_1=u\rs{p1}f\rs{o}-\frac{p}{3}\pd{}{p}\int_{-\infty}^{-0}f\frac{du}{dx}dx,
			\label{kinu:eqprom1}
		\end{equation}
		where $f\rs{o}(p)$ is the distribution function at the shock. 
		The integration of the equation (\ref{kinu:maineq}) 
		from $x=-0$ to $x=+0$ and from $x=+0$ to $+\infty$ 
		yields
		\begin{equation}
			\left[D\pd{f}{x}\right]_2-\left[D\pd{f}{x}\right]_1
			+\frac{p}{3}\pd{f\rs{o}}{p}(u_2-u_1)
			+Q\rs{o}(p)=0
			\label{kinu:eqprom2}
		\end{equation}
		\begin{equation}
			\left[D\pd{f}{x}\right]_2=
			\frac{p}{3}\pd{}{p}\int_{+0}^{+\infty}f\frac{du}{dx}dx
			-\int_{+0}^{+\infty}u\pd{f}{x}dx-u\rs{x}f\rs{o} .
			\label{kinu:eqprom3}
		\end{equation}
		Integration of the last integral in (\ref{kinu:eqprom3}) by parts transforms it to 
		\begin{equation}
			-\int_{+0}^{+\infty}u\pd{f}{x}dx=u\rs{p2}f\rs{o}.
			\label{bell2:C6}
		\end{equation}
		We introduced the notations
		\begin{equation}
			\left[D\pd{f}{x}\right]_{+\infty}=-u_3f_3\equiv -u\rs{x}f\rs{o}
			\label{bell2:C7}
		\end{equation}
		where $u_3$ and $f_3(p)$ are at the downstream infinity, $u\rs{x}$ an unknown function, 
		as well as (cf. with equations (\ref{bell2:defu1p})-(\ref{bell2:defu2p}))
		\begin{equation}
			u\rs{p1}(p)=u_1-\int_{-\infty}^{-0}\frac{f(x,p)}{f\rs{o}(p)}\frac{du}{dx}dx,
		\end{equation}
		\begin{equation}
			u\rs{p2}(p)=u_2-\int_{+\infty}^{+0}\frac{f(x,p)}{f\rs{o}(p)}\frac{du}{dx}dx
		\end{equation}
		
		With all these, the equation (\ref{kinu:eqprom2}) may be re-written as the equation for the distribution function at the shock $f\rs{o}(p)$:
		\begin{equation}
			\pd{f\rs{o}}{p}+
			\frac{f\rs{o}}{p}s\rs{o}(p)
			=\frac{3Q\rs{o}(p)}{p(u\rs{p1}-u\rs{p2})},
			\label{kinu:eqprom4}
		\end{equation}
		\begin{equation}
			s\rs{o}=\frac{3}{u\rs{p1}-u\rs{p2}}
			\left(u\rs{p1}-u\rs{p2}+u\rs{x}+\frac{p}{3}\pd{(u\rs{p1}-u\rs{p2})}{p}\right).
		\end{equation}
		The solution of this equation is 
		\begin{equation}
			f\rs{o}(p)=\frac{\eta n_1}{4\pi p\rs{o}^3}\frac{3u_1}{u\rs{p1}-u\rs{p2}}
			\exp\left[-\int_{p\rs{o}}^{p}\frac{3(u\rs{p1}-u\rs{p2}+u\rs{x})}{u\rs{p1}-u\rs{p2}}\frac{dp'}{p'}\right].
			\label{bell2:solkineq}
		\end{equation}
		The unknown function $u\rs{x}$ should be $u\rs{x}=u\rs{p2}$ in order (i) to match the expression (\ref{bell2:fnlayauongu}) derived in the frame of the approach of Bell, (ii) to give the  \citet{2002APh....16..429B} solution with  $u_2=\mathrm{const}$ \citep[cf. eq.(8) in][]{2005MNRAS.361..907B}, (iii) to yield the classic test-particle solution $f\rs{o}\propto p^{-3u_1/(u_1-u_2)}$ with $u_1=\mathrm{const}$ and $u_2=\mathrm{const}$ (then $u\rs{p1}=u_1$ and $u\rs{p2}=u_2$). 
		
		Strictly speaking, $u\rs{x}$ in the solution (\ref{bell2:solkineq}) is $u\rs{x}=u_3f_3(p)/f\rs{o}(p)$ (see eq. \ref{bell2:C7}). The equality $u\rs{x}=u\rs{p2}$ is valid in the case (i) above at least to the level of approximation for the spatial `part’ of the distribution function $f_2(x,p)$ adopted in the present paper, i.e. $f_2(x,p) \approx f\rs{o}(p){\cal H}(x\rs{p}-x)$. Indeed, let us consider some point $x\rs{p'}$ downstream. The advected flux of particles with momentum $p'$ from this point toward $x>x\rs{p'}$ is proportional to  $u_2(x\rs{p'})f\rs{o}(p')$. The flux conservation gives $u_3f_3(p')= u_2(x\rs{p'})f\rs{o}(p')$. Similarly, $u_3 f_3(p'')=u_2(x\rs{p''})f\rs{o}(p'')$ for particles with another momentum $p''$, and so on. Therefore, $u_3 f_3(p)= u_2(x\rs{p})f\rs{o}(p)$ holds for any momentum $p$ of relativistic particles. From the definition (15) and ${\cal X}_2(x) = {\cal H}(x\rs{p}-x)$, we have that $u_2(x\rs{p})=u\rs{p2}$. Thus, $u_3f_3(p)/ f\rs{o}(p) = u\rs{p2}$ and, from (\ref{bell2:C7}), $u\rs{x}=u\rs{p2}$.
			
			It would be desirable in the future to determine the distribution $f_3(p)$ from more general considerations. This task is related to the problem of distinguishing between the distribution functions of two populations of particles, namely, those which continue to participate in acceleration and those advected downstream. 
			Nevertheless, the exact evolution of particles advected toward the downstream infinity seems unimportant to determine the distribution function at the shock because they are `removed' from the accelerator. Therefore, one may set up the boundary condition (\ref{bell2:C7}) as $u_3 f_3(p)= u\rs{p2}f\rs{o}$ a priori.

	\end{appendix}
	
	
\end{document}